\documentclass[opre,nonblindrev]{informs3}

\OneAndAHalfSpacedXI

\usepackage{booktabs}
\usepackage{comment}
\usepackage{algorithm,algorithmic}
\usepackage{subcaption}
\usepackage{amsmath,amsfonts,eufrak}
\usepackage{thmtools}
\usepackage{thm-restate}
\usepackage{xcolor}
\usepackage{breqn}
\usepackage{mathtools}
\usepackage{bbm}
\usepackage{bm}
\usepackage{enumerate}
\usepackage{paralist}
\usepackage{placeins}

\usepackage{hyperref}
\hypersetup{
     colorlinks   = true,
     citecolor    = blue!55!black,
     linkcolor    = blue!55!black,
     urlcolor    = blue!55!black
}

\usepackage{soul}
\usepackage{accents}

\usepackage{xparse}

\NewDocumentCommand{\norm}{m g}{%
    \ensuremath{\left\| #1 \right\|_{\infty\IfValueT{#2}{,\, #2}}}%
}

\def\cf#1{{\sffamily \bfseries \color{blue}#1}}

\def\tcyc#1{{\sffamily \bfseries \color{red}#1}}

\def\bold#1{{\sffamily \bfseries#1}}
\newcommand\leaveline{\vspace{0.3cm}}
\newcommand\leavehalfline{\vspace{0.15cm}}
\newcommand\leavequarterline{\vspace{0.075cm}}

\def\bi{\begin{itemize}}
\def\ei{\end{itemize}}

\def\norm#1{\left\|#1\right\|}

\def\abs#1{\left|#1\right|}

\newcommand{\bE}{\mathbb{E}}

\newcommand{\bP}{\mathbb{P}}

\newcommand{\vecTtrue}{\pmb{\tau_{0}}}
\newcommand{\vecThat}{\pmb{\hat{\tau}}}

\newcommand{\vecT}{\pmb{\tau}}

\newcommand{\tauhat}{\hat{\tau}}

\newcommand{\Fhat}{\hat{F}}
\newcommand{\Ftilde}{\tilde{F}}
\newcommand{\Ftrue}{F_0}

\newcommand{\That}{\hat{t}}
\newcommand{\Ttilde}{\tilde{t}}
\newcommand{\Ttrue}{t_0}

\newcommand{\lxk}{[l,x_k]}

\graphicspath{ {images/} }

\usepackage{natbib}
 \bibpunct[, ]{(}{)}{,}{a}{}{,}%
 \def\bibfont{\small}%
\TheoremsNumberedThrough     
\ECRepeatTheorems

\EquationsNumberedThrough    

\MANUSCRIPTNO{}

\begin{document}


\RUNAUTHOR{\bold{Chan, Chen, Fernandes, and Maaz}}

\RUNTITLE{\sffamily Estimating Distributions with Failure Rate Properties from Noisy Quantile Data}

\TITLE{\Large \bold{Estimating Distributions with Failure Rate Properties \\ from Noisy Quantile Data}}

\ARTICLEAUTHORS{%
\AUTHOR{Timothy C.Y. Chan}
\AFF{Department of Mechanical and Industrial Engineering, University of Toronto}
\AUTHOR{Ningyuan Chen}
\AFF{Rotman School of Management, University of Toronto}
\AUTHOR{Craig Fernandes}
\AFF{The Wharton School, University of Pennsylvania}
\AUTHOR{Muhammad Maaz}
\AFF{Department of Mechanical and Industrial Engineering, University of Toronto}
} 

\ABSTRACT{Estimating an unknown cumulative distribution function (cdf) is fundamental both as a statistical goal and as an input to downstream optimization. In practice, however, distribution estimation is often complicated by incomplete knowledge of the distribution's structure and limited, censored data. We address the first challenge by studying distributions with \emph{failure-rate} shape constraints, especially increasing failure rate (IFR), without assuming a fully specified parametric family. To address the second, we consider \emph{noisy quantile data}: at finitely many prespecified knots, each observation records only whether an independent sample lies below or above the knot. This setting arises naturally in pricing, reliability, and healthcare. We formulate the IFR-constrained maximum likelihood estimator and show that the original problem is infinite-dimensional and non-convex. Our two-step approach solves a finite-dimensional convex optimization problem over transformed knot values, then reconstructs a full cdf through shape-preserving interpolation. We establish finite-sample error bounds and convergence rates that guide offline data collection. We also show how the framework accommodates additional properties: (i) decreasing failure rate, (ii) failure rate average, (iii) generalized failure rate, and (iv) new better than used. Numerical experiments and case studies in revenue management and reliability illustrate improved goodness of fit and downstream decision quality.
}

\KEYWORDS{Maximum likelihood estimation, current status data, shape-preserving interpolation, estimate-then-optimize}


\maketitle

\vspace{-0.75cm}

\section{Introduction} \label{sec:intro}


Estimating an unknown cumulative distribution function (cdf) from data, either as a statistical object of independent interest or as an input to a downstream optimization problem, is a fundamental problem in operations. In practice, however, distribution estimation is often complicated by (i) incomplete knowledge of the distribution's structure and (ii) limited and censored data. The following three operational settings illustrate how these challenges can arise.

\begin{example}[Revenue Management]
A monopolistic retailer seeks to estimate its customers' unknown valuation distribution to inform its pricing decisions. The retailer does not observe the customers' exact valuations. Instead, each sale or no-sale outcome reveals whether a customer's valuation exceeds the posted price. Each customer is offered only one price. For example, the data may indicate that 6 out of 10 customers purchased at the price of $\$30$, while 4 out of 8 customers purchased at the price of $\$40$.
\end{example}

\begin{example}[Reliability]
A firm operating a fleet of off-site machines seeks to estimate the unknown time-to-failure distribution to inform its maintenance decisions. Since the machines are inspected at scheduled ages, their exact failure times are unavailable. Each inspection reveals whether a machine has failed or remains operational at that time. Each machine is inspected only once, at a prespecified age. For example, one may observe that 3 out of 5 machines remain operational at a 6-month inspection.
\end{example}

\begin{example}[Healthcare]
A drug manufacturer seeks to estimate the unknown distribution of patients' minimum-effective-dosage to inform treatment decisions. Rather than observing each patient's exact minimum effective dosage, researchers conduct experiments at a finite set of dosage levels. Each observation reveals whether a patient responds favorably at the administered level. Each patient receives only one dosage level. For example, a clinical trial may test 50 patients at 10 mg and 50 patients at 20 mg, with 25 patients responding favorably at the lower dosage and 35 responding favorably at the higher dosage.
\end{example}

Despite their different applications, these examples share two defining features. First, practitioners may lack a reliable parametric specification for the ground-truth distribution, yet domain knowledge may support global shape restrictions. In this paper, we focus on distributions satisfying one of many possible \emph{failure-rate} properties. Perhaps the most prevalent example is the \emph{increasing failure rate} (IFR) property, which is widely used in reliability, economics, and operations applications.\footnote{In other fields, the IFR property is often called the increasing hazard rate property.}  At a high level, the IFR property means that, conditional on a sample exceeding a threshold, the probability that it lies within the next fixed interval does not decrease as the threshold rises. The class is broad, including uniform, normal, exponential, logistic, extreme-value, Laplace, Rayleigh, and Maxwell distributions, as well as power-function, Weibull, gamma, chi-square, chi, and beta distributions under appropriate parameter restrictions \citep{bagnoli2005}. Returning to the above examples, customer valuation, machine time-to-failure, and minimum-effective-dosage distributions are each commonly modeled as IFR \citep{bagnoli2005, barlow1996mathematical, schell1989increasing}.

The second feature is that the data do not consist of exact realizations from the distribution, as is commonly assumed in many theoretical studies of distribution estimation. Rather, in each example above, we assume that the underlying customer valuations, machine failure times, or patients' minimum effective dosages are iid, with each observation only recording a binary comparison at one of a small number of prespecified values. We refer to these values, or thresholds, as \emph{knots}. In the three examples above, the knots correspond to prices \citep{allouah2023singlepoint, bahamou2026fast}, inspection times \citep{jardine2013maintenance}, and dosage levels \citep{chen2024adaptive, manski2025using}, respectively. Aggregating multiple binary observations collected at each knot yields a noisy estimate of the cdf value (equivalently, the quantile value) for that knot. We refer to this data regime as \emph{noisy quantile data}. It is closely related to case I interval censoring, also known as current status data \citep{wellner1992interval}.

Together, these two practical features create a challenging and understudied estimation problem: reconstructing an entire distribution from noisy observations at finitely many points while preserving a global shape restriction. Accordingly, our goal is to \emph{estimate distributions with failure-rate shape constraints from noisy quantile data}. Although prior work has studied other shape constraints under this data regime, to our knowledge, this is the first to consider the failure-rate properties examined here. We develop a tractable estimation framework with theoretical guarantees and demonstrate its value across several practical domains. Our main contributions are: 

\leaveline
\begin{enumerate}

    \item \emph{Estimation Framework.} We formulate the problem of fitting an IFR-constrained distribution from noisy quantile data as a maximum likelihood estimation problem. The estimator is a solution to an infinite-dimensional non-convex optimization problem, making direct computation impractical (Section \ref{sec:problem}). We develop a tractable approach that computes an optimal solution by (i) solving a finite-dimensional convex problem over the knots and (ii) applying a shape-preserving interpolation method to recover the full distribution 
    (Section \ref{sec:solution-framework}). 
    %

    \item \emph{Theoretical Guarantees.} 
    We provide a finite-sample bound on the error between our estimated cdf and the ground-truth cdf, and establish the asymptotic convergence rate (Section \ref{sec:theory}). 
    These results also yield prescriptive operational guidelines for offline data collection. 

    \item \emph{Additional Failure-Rate Properties.} Leveraging a similar reformulation technique as in the IFR case, we rigorously discuss how our framework can flexibly capture other failure rate properties including (i) decreasing failure rate, (ii) failure rate average, (iii) generalized failure rate, and (iv) new better than used, highlighting the generalizability of our approach (Section \ref{sec:extensions}). 

    \item \emph{Numerical Experiments and Case Studies.} We conduct numerical experiments to empirically (i) validate the theoretical guidance for offline data collection, and (ii) demonstrate that our estimator outperforms benchmark alternatives in terms of goodness-of-fit and computation time (Section \ref{sec:numerics}). Finally, we present case studies in revenue management and reliability. We provide theoretical guarantees of decision quality and numerically show that our estimator improves downstream decisions in realistic estimate-then-optimize settings (Section \ref{sec:case-study}).

    \item \emph{Software.} We provide open-source Python code for fitting IFR distributions and the four additional failure-rate properties, and for reproducing all experiments and case studies.\footnote{ The code is available here: 
    \href{https://github.com/craigfernandes1/Fitting-IFR-CDFs-from-Noisy-Quantile-Data}{github.com/craigfernandes1/Fitting-IFR-CDFs-from-Noisy-Quantile-Data}.
    }
\end{enumerate}

\subsection{Related Literature} \label{sec:lit}

Our work develops a method to estimate failure-rate constrained distributions from noisy quantile data, which may be used in downstream optimization models. It thus connects to two broad streams of literature: (i) estimating shape-constrained distributions and (ii) data-driven optimization.

\bold{Estimating Shape-Constrained Distributions.} A substantial body of work studies the estimation of an unknown cdf under structural constraints. Much of this literature assumes access to direct samples from the underlying distribution, possibly subject to censoring. When the distribution belongs to a known parametric family, classical approaches such as maximum likelihood estimation or the method of moments are typically used \citep{casella2002statistical}. In nonparametric settings, common approaches include fitting the empirical cdf or applying kernel density estimation \citep{tsybakov2009introduction}. When additional structural information (e.g., shape constraints) is available, specialized techniques have been developed. Of particular relevance to our work, early studies by \citet{grenander1956}, \citet{marshall1965}, and \citet{prakasa1970} proposed and analyzed estimators for distributions with monotone failure rates using full sample data.\footnote{This stream of work is also broadly related to studies on estimating general shape-constrained functions, rather than distributions specifically, including single-variable functions \citep{guntuboyina2018nonparametric} and multivariate functions \citep{lim2012consistency, bertsimas2020sparse,lin2022augmented}.}  In the context of censored observations from these types of distributions, \citet{padgett1980} developed an estimator based on tools from the censored data literature and compared their method to the well-studied Kaplan--Meier estimator \citep{kaplan1958}. 

Closer to our data regime, several works have developed nonparametric estimators given interval-censored data. In this setting, rather than observing an exact realization from the distribution, one observes only an interval known to contain it. A special case is \emph{case I interval censoring}, or \emph{current status data}, in which each observation records only whether a sample lies below or above a known inspection threshold. Maximum likelihood estimators have been proposed in this setting, with the resulting cdf estimate typically taking the form of an isotonic-regression-type step function \citep{ayer1955empirical, Turnbull1976, wellner1992interval, huang1997interval, sun2006intervalcensored, groeneboom2014nonparametric}. 

Closest to our work are studies of shape-constrained estimation from interval-censored data, primarily motivated by survival analysis.  \citet{dumbgen2004consistency, dumbgen2006estimating} consider concave cdfs, while \citet{dumbgen2014maximum} and \citet{andersonbergman2016computing} consider log-concave densities. Finally, in parallel development with our work, \citet{chu2026nonparametric} generalize these shape restrictions by studying log-concave cdfs under mixed-case interval censoring, which includes case I observations. Our focus differs: \cite{chu2026nonparametric} analyze random inspection times and allow repeated inspections of a subject, whereas we study independent binary observations at a finite set of prespecified knots, with sample sizes allocated across them in advance. We begin by imposing an IFR shape constraint\footnote{Our baseline IFR constraint is equivalent, under reflection, to log-concavity of the cdf. The additional failure-rate properties we study are distinct from those studied in \cite{chu2026nonparametric}.}, and use a transformed-space convex formulation analogous to theirs. Our contributions differ in three respects. First, Section \ref{sec:solution-framework} establishes the existence of an optimal IFR cdf for our fixed-knot problem. Section \ref{sec:theory} then provides finite-sample
error bounds over the entire known support, separating error at the knots and between the knots to identify how sampling noise and interpolation error impact goodness of fit. By comparison, \citet{chu2026nonparametric} establish asymptotic Hellinger convergence rates for the observed-data distribution under specific technical conditions, and do not provide finite-sample bounds. Second, Section \ref{sec:extensions} demonstrates the robustness and generality of our framework. With suitable modifications to our main algorithm, we can estimate distributions satisfying decreasing failure rate, increasing failure-rate average, generalized failure rate, and new better than used properties, none of which are addressed by the preceding interval-censoring studies. Third, we take an operations perspective by providing theory-informed guidance on how to choose the number and locations of knots, as well as how to allocate observations across them. We also evaluate the fitted cdf as an input to downstream pricing and preventive-maintenance decisions and provide guarantees on the resulting decision quality.

\bold{Data-Driven Optimization.} Our work also relates to the literature on using data to estimate an input model for a downstream optimization problem. In contrast to decision-specific approaches, our goal is to estimate a full distribution that can be reused across operational decisions, while only considering noisy quantile data rather than exact samples from the ground-truth distribution. This places our work within the ``estimate, then optimize'' framework, in which a distributional object is first learned from data and then used as an input to decision-making. Note that this is distinct from the ``predict, then optimize'' framework which focuses on fitting point estimates rather than an entire distribution. 

Notable examples of the estimate, then optimize framework include estimating the distribution of travel times on a road network to optimize route selection, or estimating rainfall distributions to support harvest planning. Most relevant to our study are applications in revenue management (see \citet{chen2023datadriven} for a review), where researchers have explored estimating demand or consumer valuation distributions to inform optimal pricing and inventory decisions (e.g., \citet{caro2012clearance, ferreira2015analytics, besbes2020rotable, boada2019inventory, arslan2021sports}). These works have considered both parametric and nonparametric models.

More recently, an emerging paradigm known as ``end-to-end optimization'' has sought to bypass explicit distribution fitting, instead learning decision policies directly from data (e.g., \citet{wilder2019decisionfocused, mandi2022decisionfocused, ban2019bigdata, elmachtoub2021smart, bertsimas2019prescriptive, abouee2022datadriven, babier2024learning}). See also \citet{chen2023datadriven} for an overview of this trend within revenue management. In the context of pricing, a line of research has explored optimal pricing given a finite number of samples from the unknown distribution \citep{fu2015randomization, huang2015samples, daskalakis2020sdp,babaioff2018two, allouah2022pricing}. More recently, several works suggest robust prices given  noiseless quantiles of the distribution  \citep{allouah2023singlepoint, daei2025robust, bahamou2026fast}. In contrast to these approaches, we position our work within the estimate, then optimize framework, as our estimation method is not tied to a particular application domain. Moreover, we consider the practical setting where the decision-maker observes only estimates of the cdf at a limited number of knots.

Finally, our work relates to the sample complexity of estimating cdfs. For example, \citet{leme2023description} develop an adaptive procedure using binary pricing queries to approximate an unknown regular cdf in Lévy distance. Although their analysis exploits regularity, the resulting approximation need not preserve regularity or satisfy IFR. In contrast, our estimated cdf satisfies IFR, which is important for downstream optimization.

\bold{Our paper.} In summary, our work estimates distributions with previously unstudied failure-rate properties under a special case of current status data (i.e., finite knots and grouped observations). We provide an estimation framework with finite sample bounds and convergence rates that can be applied across a range of downstream operational decision tasks.

\section{Problem Definition} \label{sec:problem}




Consider a random variable $X$ with unknown cdf $F_0$ and known full support $[l,u]$. For ease of exposition, we assume throughout that both $l$ and $u$ are finite; Remark \ref{rem:unbounded} discusses how the framework naturally extends to unbounded supports. 
We assume that $F_0$ is absolutely continuous, with unknown density $f_0$. 
Rather than observing realizations of $X$ directly, we observe only whether independent draws fall below a threshold. Specifically, let $l<x_1<\cdots<x_k<u$ denote $k$ predefined thresholds, or ``knots.'' A single observation at knot $x_i$ is a Bernoulli trial with success probability $F_0(x_i)$, where success corresponds to the event $X \le x_i$. At each knot $x_i$, we observe $n_i\ge1$ independent trials and record $y_i$ successes. Thus, $y_i \sim \mathrm{Binomial}(n_i,F_0(x_i))$, independently across $i=1,\ldots,k$. Equivalently, $y_i/n_i$ is an estimate of $F_0(x_i)$, i.e., what we call a \emph{noisy quantile}.

We focus on fitting the underlying cdf itself because downstream decisions may vary across applications and no single decision-specific objective applies to all settings. Given our noisy quantile data, we choose $F$ to maximize the binomial log-likelihood, whose expected value is highest when $F$ agrees with $F_0$ at every observed knot:
\begin{align*}
    \sum_{i=1}^k \Big( y_i \ln F(x_i) + (n_i - y_i) \ln (1-F(x_i)) \Big).
\end{align*}
We adopt the conventions $\log(0)=-\infty$ and $0\log(0)=0$. Additionally, we assume that $\Ftrue$ is IFR. Accordingly, in addition to the constraints ensuring that $F$ is a valid cdf, we impose the IFR shape constraint on $F$; alternate shape constraints are rigorously discussed in Section \ref{sec:extensions}.
%
Mathematically, a distribution with cdf $F$ and density $f$ is IFR if its failure rate (or hazard rate) $h(x)$ is non-decreasing on $[l,u)$, where 
\begin{align*}
    h(x) = \frac{f(x)}{1-F(x)}.
\end{align*}
%
An equivalent definition of IFR, which we primarily use, is that the survival function $1-F(x)$ is log-concave \citep{barlow1996mathematical}.
%
Putting this together, our optimization problem of interest is: 
%
\begin{subequations}
\label{eqn:originalprob}
\begin{align}
    \sup_{F} \quad & \sum_{i=1}^k \Big( y_i \ln F(x_i) + (n_i - y_i) \ln (1-F(x_i)) \Big) \label{eq:log-LL} \\
    \text{s.t.} \quad  & 0 \le F(x) \leq 1 \quad \forall x \in [l,u], \label{eq:1b} \\
    & F(l) = 0, \:\: F(u) = 1,  \label{eq:1c} \\
    & F(x) \text{ is absolutely continuous and non-decreasing on } [l,u], \label{eq:1d}  \\
    & 1-F(x) \text{ is log-concave on } [l,u]. \label{eq:origprob-ifr-constraint}
\end{align}
\end{subequations}

Problem \eqref{eqn:originalprob} is \emph{infinite-dimensional} and \emph{non-convex}. To establish the latter, the following example demonstrates that convex combinations of IFR distributions need not be IFR.


\begin{example}[\bold{Non-convexity}]
    Consider the support $[0,1]$ and the two functions $F_1(x)=(1-e^{-x})/(1-e^{-1})$ and $F_2(x)=(1-e^{-5x})/(1-e^{-5})$. Both satisfy conditions \eqref{eq:1b}--\eqref{eq:1d}, making them valid cdfs. Moreover, direct differentiation shows that $\ln(1-F_j(x))$ is concave for $j=1,2$, and hence both distributions are IFR. Now, let $F_3(x)=(F_1(x)+F_2(x))/2$. Although $F_3$ remains a valid cdf, it is not IFR: evaluating the second derivative of $\ln(1-F_3(x))$ at $x=0$ gives approximately $2.433>0$. Thus, $\ln(1-F_3(x))$ is not concave, implying 
    the feasible region of \eqref{eqn:originalprob} is non-convex.
\end{example}



One way to deal with the infinite-dimensional nature of \eqref{eqn:originalprob} is to discretize the support, an approach we use as a benchmark in Section \ref{sec:numerics}.  However, discretization alone does not remove the computational difficulty induced by the IFR constraint. To see this, consider an equally spaced grid and any three consecutive points $z_1 < z_2 < z_3$. A natural discrete analogue of the IFR condition is
\begin{align*}
\frac{F(z_2) - F(z_1)}{1 - F(z_1)} \leq \frac{F(z_3) - F(z_2)}{1 - F(z_2)}
\iff (1 - F(z_2))^2 \geq (1 - F(z_1))(1 - F(z_3)).
\end{align*}
Thus, even after discretization, the IFR constraint introduces non-convex quadratic inequalities. Since these constraints must be imposed across consecutive triplets of grid points, a direct discretization leads to a non-convex quadratically constrained likelihood-maximization problem that may be computationally prohibitive.

\section{A Tractable Solution Framework} \label{sec:solution-framework}

In this section, we develop a two-step approach for solving \eqref{eqn:originalprob}. First, we use a change of variables to reformulate the infinite-dimensional non-convex problem as a finite-dimensional convex program over the knots. Second, we recover a full distributional estimate by interpolating between the optimized knot values. Let 
\begin{align*}
t(x) = \ln(1-F(x)),
\end{align*}
so that $F(x)=1-e^{t(x)}$.
The log-likelihood becomes
\begin{align*}
\sum_{i=1}^k \Big(y_i \ln(1-e^{t(x_i)}) + (n_i-y_i)t(x_i)\Big).
\end{align*}
The range, endpoint, and monotonicity constraints translate naturally into $t$-space: (i) $0\le F(x)\le 1$ becomes $t(x)\le 0$ for all $x\in[l,u)$, (ii) $F(l)=0$ and $F(u)=1$ become $t(l)=0$ and $t(x)\to -\infty$ as $x\to u$, and (iii) $F(x)$ being non-decreasing becomes $t(x)$ being non-increasing. Absolute continuity is not encoded by the finite knot values alone; it is ensured by the continuous reconstructions introduced below.  Finally, the IFR constraint \eqref{eq:origprob-ifr-constraint} admits a convenient reformulation: $1-F(x)$ is log-concave if and only if $t(x)$ is concave. 

Next, we exploit the fact that the likelihood depends only on the values of $t(\cdot)$ at the knots. Thus, rather than optimizing over the full function $t(\cdot)$, we solve for the likelihood-maximizing knot values. Let $x_0=l$, and define the knot-value vector
\begin{align*}
\vecT := (\tau_0,\tau_1,\ldots,\tau_k)\in\mathbb{R}^{k+1},
\end{align*}
%
%
where $\tau_i = t(x_i)$ for $i=0,\ldots,k$, i.e., the value of $t(x)$ at knot $x_i$. The monotonicity and concavity requirements on $t(\cdot)$ are then imposed on $\vecT$, yielding the following finite-dimensional formulation:
\begin{subequations}
\label{eqn:covprob}
\begin{align}
\sup_{\vecT} \quad & \sum_{i=1}^k \Big( y_i \ln(1-e^{\tau_i}) + (n_i-y_i)\tau_i \Big) \\
\text{s.t.} \quad & \tau_i \le 0 \quad \forall i=1,\ldots,k, \\
& \tau_0 = 0, \\
& \tau_i \geq \tau_{i+1} \quad \forall i=0,\ldots,k-1, \\
& \frac{\tau_i-\tau_{i-1}}{x_i-x_{i-1}} \geq \frac{\tau_{i+1}-\tau_i}{x_{i+1}-x_i} \quad \forall i=1,\ldots,k-1. \label{eq:covprob-ifr-constraint}
\end{align}
\end{subequations}
%
%
In addition to being finite-dimensional, \eqref{eqn:covprob} is convex.

\begin{restatable}[\bold{Convexity}]{lemma}{ConvexProblem}
\label{lem:convex-problem}
    Problem \eqref{eqn:covprob} is a convex optimization problem. 
\end{restatable}

Thus, \eqref{eqn:covprob} can be handled using standard convex optimization methods whenever its supremum is attained. Because its feasible region is unbounded, however, attainment and uniqueness require additional care. The following lemma characterizes both properties.

\begin{restatable}[\bold{Existence and Uniqueness}]{lemma}{ConvexOptUnique}
\label{lem:opt_unique}
    An optimal solution to \eqref{eqn:covprob} exists if and only if $y_k<n_k$. When $y_k<n_k$, the optimal knot vector is unique if $y_i>0$ for all $i=1,\ldots,k$. If $y_i=0$ at one or more knots, uniqueness is not guaranteed.
\end{restatable}


The condition $y_k<n_k$ is necessary and sufficient for the existence of a finite optimal knot vector. When $y_k=n_k$, the likelihood improves as $\tau_k\to-\infty$, and the supremum in \eqref{eqn:covprob} is not attained. In contrast, an outcome of $y_i=0$ at any knot does not affect existence, but it may destroy uniqueness because the corresponding transformed log-likelihood contribution is linear rather than strictly concave. Thus, $y_i>0$ for all $i=1,\ldots,k$ is sufficient, but not necessary, for uniqueness. For fixed $k$, the events $y_k=n_k$ and $y_i=0$ have vanishing probability as $\min_{i=1,\ldots,k}n_i\to\infty$. 
Moreover, although \eqref{eqn:covprob} has fewer decision variables than \eqref{eqn:originalprob}, the following lemma shows that the two problems have equal objective values.

\begin{restatable}[\bold{Equivalence}]{lemma}{KnotValueEquivalence}
\label{lem:knot_value_equivalence}
    The objective values of \eqref{eqn:originalprob} and \eqref{eqn:covprob} are equal.
\end{restatable}


Since \eqref{eqn:covprob} determines only the knot values $\vecT$, obtaining a full estimator requires a reconstruction of $t$ that matches these values, remains globally non-increasing and concave, and satisfies the endpoint condition. We obtain such a reconstruction by using piecewise-linear interpolation in $t$-space between the fitted knot values and appending an analytic tail that preserves these properties. The full procedure is summarized in Algorithm \ref{alg:main}. Theorem \ref{thm:optimal-solution} establishes that Algorithm \ref{alg:main} returns an optimal solution to the original infinite-dimensional problem.

\begin{algorithm}
\caption{IFR Distribution Estimator with Noisy Quantile Data}
\label{alg:main}
\begin{algorithmic}[1]
\REQUIRE Knots $x_1,\ldots,x_k$ with $n_i$ samples and $y_i$ observations of $X\leq x_i$, for $i=1,\ldots,k$.
\ENSURE A cdf $\Fhat$ satisfying IFR and maximizing the log-likelihood \eqref{eq:log-LL}.
\IF{$y_k<n_k$}
\STATE Solve \eqref{eqn:covprob} to obtain an optimal solution $\vecThat=(\tauhat_0,\tauhat_1,\ldots,\tauhat_k)$.
\STATE Construct $\That$ by piecewise-linear interpolation through $(x_i,\tauhat_i)$ for $i=0,\ldots,k$.
\STATE Choose $a>0$ such that $a\geq -(\hat{\tau}_k-\tauhat_{k-1})(u-x_k)/(x_k-x_{k-1})$.
\STATE Define $\That(x):=\tauhat_k+a\log\left(\frac{u-x}{u-x_k}\right)$ for $x\in[x_k,u)$.
\ELSE
\STATE Let $j:=\max\{i\in\{1,\ldots,k-1\}:y_i<n_i\}$, with $j:=0$ if this set is empty.
\IF{$j\geq1$}
\STATE Solve \eqref{eqn:covprob} only on the knots $x_1,\ldots,x_j$ to obtain an optimal solution $(\tauhat_0,\tauhat_1,\ldots,\tauhat_j)$.
\STATE Construct $\That$ by piecewise-linear interpolation through $(x_i,\tauhat_i)$ for $i=0,\ldots,j$.
\STATE Choose $a>0$ such that $a\geq - (\tauhat_j-\tauhat_{j-1})(x_{j+1}-x_j)/(x_j-x_{j-1})$.
\ELSE
\STATE Set $\tauhat_0:=0$ and choose any $a>0$.
\ENDIF
\STATE Define $\That(x):=\tauhat_j+a\log\left(\frac{x_{j+1}-x}{x_{j+1}-x_j}\right)$ for $x\in[x_j,x_{j+1})$ and $\That(x):=-\infty$ for $x\in[x_{j+1},u]$.
\ENDIF
\STATE Define $\Fhat(x):=1-\exp(\That(x))$ for $x\in[l,u)$, with the convention $\exp(-\infty)=0$; and $\Fhat(u):=1$.
\RETURN $\Fhat$.
\end{algorithmic}
\end{algorithm}

\begin{restatable}[\bold{Optimal Solution}]{theorem}{ReconstructionCoincides}
\label{thm:optimal-solution}
Problem \eqref{eqn:originalprob} admits an optimal solution for every realization of the data. Moreover, every cdf $\Fhat$ returned by Algorithm \ref{alg:main} is an optimal solution to \eqref{eqn:originalprob}. 
\end{restatable}

\begin{figure}[tbh]
    \centering
    \includegraphics[width=0.65\linewidth]{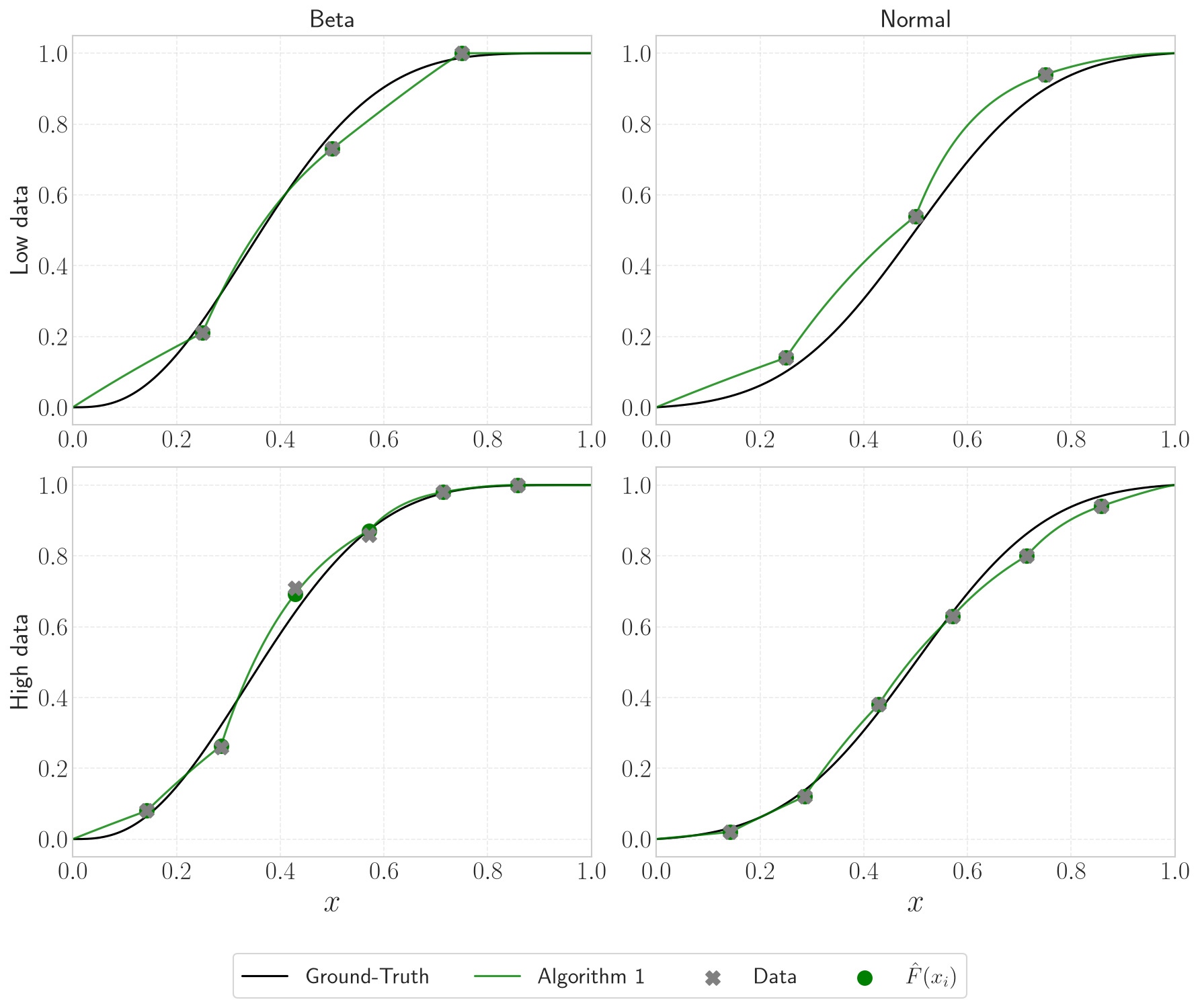}
    \caption{\normalfont Numerical example comparing $\Fhat$ from Algorithm \ref{alg:main} against two IFR ground-truth cdfs supported on $[0,1]$. (Left) $\Ftrue\sim\text{Beta}(3,5)$. (Right) $\Ftrue\sim \text{Normal}(0.5, 0.2)$. (Top) $k=3$ equidistant knots. (Bottom) $k=6$ equidistant knots. Each knot has $n_i=100$ samples.}
    \label{fig:visualization}
\end{figure}

Thus, an optimal solution to \eqref{eqn:originalprob} always exists, and Algorithm \ref{alg:main} returns one such solution.\footnote{Within Algorithm \ref{alg:main}, when $y_k<n_k$, our implementation sets $a:=\max\left\{1,-(u-x_k)(\tauhat_k-\tauhat_{k-1})/(x_k-x_{k-1})\right\}$. When $y_k=n_k$ and $j\geq1$, it sets $a:=\max\left\{1,-(x_{j+1}-x_j)(\tauhat_j-\tauhat_{j-1})/(x_j-x_{j-1})\right\}$; when $j=0$, it sets $a:=1$.} However, even when \eqref{eqn:covprob} has a unique optimal knot vector, the full cdf solving \eqref{eqn:originalprob} need not be unique. Because the likelihood depends only on the values at the observed knots, it generally does not identify the cdf between knots.
The behavior of Algorithm~\ref{alg:main} across two ground-truth distributions and two data regimes is illustrated in Figure~\ref{fig:visualization}.  



\leaveline
\begin{remark}[\bold{Unbounded Supports}]
\label{rem:unbounded}
For expositional convenience, Algorithm \ref{alg:main} is stated for finite endpoints $l$ and $u$. The construction extends to unbounded supports by appending suitable non-increasing, concave tails in $t$-space. When $l=-\infty$, replace the artificial lower knot $x_0=l$ and its associated constraints with a compatible lower-tail construction that matches the interpolant at the first observed knot and satisfies $\That(x)\to0$ as $x\to-\infty$.\footnote{If $\hat\tau_1=\hat\tau_2<0$, no compatible lower tail approaches $0$ as $x\to-\infty$. A slight perturbation of the fitted knots yields a continuous IFR cdf with arbitrarily small log-likelihood loss; see Section \ref{sec:extensions} for related endpoint cases.} When $u=\infty$, append an upper tail that matches the interpolant at the last observed knot and satisfies $\That(x)\to-\infty$ as $x\to\infty$. In each case, the tail is chosen to preserve concavity at the joining knot. Consequently, the resulting cdf satisfies $\Fhat(x)\to0$ as $x\to-\infty$ and $\Fhat(x)\to1$ as $x\to\infty$. The case in which $y_k = n_k$ is unchanged, since it sets $\Fhat(x)=1$ after the first terminal all-success knot. \hfill \Halmos
\end{remark}

\begin{remark}[\bold{Beyond Piecewise Linear}]
\label{rem:beyond-pwl}
We highlight that Algorithm~\ref{alg:main} interpolates the fitted knot values piecewise linearly in \(t\)-space, while the estimated cdf \(\Fhat(x)=1-\exp(\That(x))\) is generally nonlinear between knots. It may have kinks at the knots, but remains absolutely continuous and has a density almost everywhere. We focus on this reconstruction because it is simple to implement, performs well in Sections~\ref{sec:numerics} and~\ref{sec:case-study}, and permits the reconstruction-specific analysis in Section~\ref{sec:theory}. Other non-increasing, concave interpolators, such as the quadratic spline of \cite{schumaker1983shape}, may also be used, but choosing their slope coefficients and appending a compatible tail requires additional care to preserve global concavity. The general finite-sample bound and consistency result of Corollary~\ref{cor:baseline_error_f} also hold for any valid interpolator, including a suitable quadratic spline. \hfill \Halmos
\end{remark}

\section{Theoretical Guarantees} \label{sec:theory}

In this section, we analyze and bound the error\footnote{We consider the sup-norm between the estimated and true cdfs, i.e., the Kolmogorov–Smirnov distance metric.} between the output from Algorithm~\ref{alg:main}, $\Fhat$, and the ground truth, $\Ftrue$. Recall that the algorithm proceeds in two conceptually distinct steps. First, it estimates the cdf values at the observed knots through the finite-dimensional likelihood-maximization step of Algorithm~\ref{alg:main}. Second, conditional on those fitted knot values, it deterministically reconstructs a cdf over the full support via interpolation and extrapolation. We therefore first study the stochastic estimation error at the knots and derive a reconstruction-agnostic full-support guarantee. We then specialize to piecewise-linear interpolation and quantify the error introduced by this particular reconstruction. Finally, we use these results to provide guidance on offline data collection under a fixed sampling budget.

For notational convenience, all norms in this section are understood to be sup-norms, and we therefore do not include the ``$\infty$'' subscript. For a function $t$ and interval $[a,b]$, we define $\norm{t}_{[a,b]}:=\sup_{x\in[a,b]}\abs{t(x)}$. Let $x_{k+1}:=u$, $N:=\sum_{i=1}^k n_i$, $\underline n:=\min_{i=1,\ldots,k}n_i$, and $\Delta:=\max_{i=0,\ldots,k-1}(x_{i+1}-x_i)$.

\subsection{Knotwise Estimation Error}
\label{sec:stochastic-estimation-error}

We first study the error in the fitted cdf values at the observed knots. This error comes entirely from the finite-dimensional fitting step of Algorithm~\ref{alg:main} and therefore does not depend on the subsequent reconstruction between knots.

\begin{restatable}[\bold{Knotwise Estimation Error}]{proposition}{EstimationErrorFspace}
\label{prop:estimation_error_f}
Let $\Fhat$ be a cdf returned by Algorithm \ref{alg:main}. For every $\delta\in(0,1)$, with probability at least $1-\delta$,
\begin{align*}
\max_{i=1,\ldots,k}\abs{\Fhat(x_i)-\Ftrue(x_i)}\le\sqrt{\frac{k\ln 3+\ln(2/\delta)}{\underline n}}+\sqrt{\frac{\ln(4k/\delta)}{2\underline n}}.
\end{align*}
\end{restatable}

Proposition~\ref{prop:estimation_error_f} controls the fitted cdf values at the observed knots. The first term bounds the adjustment from the empirical cdf values induced by likelihood maximization under the IFR constraint, while the second is the usual binomial sampling error. The IFR constraints couple the fitted values across knots, so improving the overall likelihood can increase the estimation error at an individual knot. Consequently, the binomial concentration bound for the empirical proportions does not transfer directly to the fitted values. Our analysis controls this adjustment through the aggregate likelihood loss.
This knotwise bound immediately yields a general guarantee over the full support. The result requires only that the reconstruction interpolate the fitted knot values and produce a valid cdf; it does not otherwise depend on the particular interpolation scheme. Notably, this also implies that the consistency result remains valid under mild modifications of the more complicated shape classes considered in Section \ref{sec:extensions}.

\begin{restatable}[\bold{General Error Bound}]{corollary}{BaselineErrorFspace}
\label{cor:baseline_error_f}
Let $\Fhat$ be a cdf returned by Algorithm \ref{alg:main}. For every $\delta\in(0,1)$, with probability at least $1-\delta$,
\begin{align*}
\norm{\Fhat-\Ftrue}_{[l,u]}\le\sqrt{\frac{k\ln 3+\ln(2/\delta)}{\underline n}}+\sqrt{\frac{\ln(4k/\delta)}{2\underline n}}+\max_{i=0,\ldots,k}\left\{\Ftrue(x_{i+1})-\Ftrue(x_i)\right\}.
\end{align*}
If $\norm{f_0}_{[l,u]}<\infty$, then
\begin{align*}
\norm{\Fhat-\Ftrue}_{[l,u]}\le\sqrt{\frac{k\ln 3+\ln(2/\delta)}{\underline n}}+\sqrt{\frac{\ln(4k/\delta)}{2\underline n}}+\max\left\{\norm{f_0}_{[l,u]}\Delta,1-\Ftrue(x_k)\right\}.
\end{align*}
\end{restatable}

Corollary~\ref{cor:baseline_error_f} controls the entire reconstruction using only cdf monotonicity. Its deterministic term is the largest true cdf increment between consecutive knots. Despite the seeming simplicity of this bound, Section~\ref{sec:tightness-example} gives a cdf and data realization for which the realized error is 98.7\% of the bound, highlighting its near-tightness.
Moreover, this corollary implies uniform consistency whenever $\Delta\to0$, $x_k\to u$, and $k/\underline n\to0$. Under equal allocation and equidistant placement, $\underline n=N/k$ and $\Delta=(u-l)/(k+1)=u-x_k$. If, in addition, $f_0$ is bounded, then the upper bound has order
\begin{align*}
\norm{\Fhat-\Ftrue}_{[l,u]}=O_p\left(\frac{k}{\sqrt{N}}+\frac{1}{k}\right).
\end{align*}

\subsection{Reconstruction Error}
\label{sec:deterministic-reconstruction-error}

We now specialize to piecewise-linear interpolation in the transformed $t$-space to derive an error bound specific to Algorithm \ref{alg:main}. Let $\vecTtrue:=(\Ttrue(x_0),\Ttrue(x_1),\ldots,\Ttrue(x_k))$ denote the true transformed knot values, e.g., $t_0(x_i) = \ln(1-F_0(x_i))$. Let $\Ttilde$ be the piecewise-linear interpolant through $\vecTtrue$, and define $\Ftilde(x):=1-\exp(\Ttilde(x))$ on $\lxk$. Thus, $\Ftilde$ is the oracle piecewise-linear reconstruction obtained from the true transformed knot values. Define
\begin{align*}
R:=\max_{i=0,\ldots,k-1}\frac{1-\Ftrue(x_i)}{1-\Ftrue(x_{i+1})}.
\end{align*}
Because $x_k<u$ and $\Ftrue$ has full support $[l,u]$, every denominator is positive. Moreover, $R\ge1$ because $\Ftrue$ is non-decreasing. We refer to $R$ as the \emph{survival decay ratio}: it measures the largest proportional decrease in survival across adjacent observed knots.

On the observed region, we decompose the error as
\begin{align*}
\norm{\Fhat-\Ftrue}_{\lxk}\le\norm{\Fhat-\Ftilde}_{\lxk}+\norm{\Ftilde-\Ftrue}_{\lxk}.
\end{align*}
The first term captures the effect of reconstructing from the fitted cdf values at the knots rather than the true knot values, while the second captures the error that remains even with oracle access to the true knot values.

\begin{restatable}[\bold{Reconstruction Error}]{lemma}{ReconstructionStability}
\label{lem:reconstruction_stability}
Let $\Fhat$ be a cdf returned by Algorithm~\ref{alg:main}. Then,
\begin{align*}
\norm{\Fhat-\Ftilde}_{\lxk}\le R\max_{i=1,\ldots,k}\abs{\Fhat(x_i)-\Ftrue(x_i)}.
\end{align*}
\end{restatable}

Lemma~\ref{lem:reconstruction_stability} holds for every realization of the data, including the terminal-all-success branch of Algorithm~\ref{alg:main}, in which $\Fhat$ is not piecewise-linear in $t$-space on the final observed cells. It shows that reconstruction amplifies knotwise cdf error by at most $R$, so Proposition~\ref{prop:estimation_error_f} bounds this term with high probability. We next bound the second term.

\begin{restatable}[\bold{Interpolation Error}]{lemma}{InterpolationErrorFspace}
\label{lem:interpolation_error_f}
Suppose that $\Ttrue\in C^2\lxk$\footnote{Here, $C^2[l,x_k]$ denotes the set of functions that are twice continuously differentiable on $[l,x_k]$.}. Then, 
$\Ftilde(x)\ge\Ftrue(x)$ for every $x\in\lxk$, and
\begin{align*}
\norm{\Ftilde-\Ftrue}_{\lxk}\le\frac{R\Delta^2}{8}\norm{(1-\Ftrue)\Ttrue''}_{\lxk}.
\end{align*}
\end{restatable}
Lemma~\ref{lem:interpolation_error_f} quantifies how the accuracy of the oracle piecewise-linear reconstruction depends on the grid and the local shape of the ground-truth distribution. The $t$-space piecewise-linear interpolation remainder (i.e., $\Delta^2/8$) is standard in the numerical analysis literature \citep[Chapter 11]{ascher2011first}; our analysis propagates this remainder through the nonlinear mapping $F(x)=1-e^{t(x)}$ and expresses the resulting cdf error in terms of survival-weighted curvature. In particular, the displayed bound is of second order in $\Delta$ whenever $R\norm{(1-\Ftrue)\Ttrue''}_{\lxk}=O(1)$ as the grid is refined.
Lemma~\ref{lem:interpolation_error_f} also shows that $\Ftilde(x)\ge\Ftrue(x)$ throughout the observed region, so the oracle reconstruction never underestimates the true cdf. Finally, the smoothness condition has a familiar interpretation: because $1-\Ftrue$ is strictly positive on $\lxk$, $\Ttrue\in C^2\lxk$ is equivalent to $\Ftrue\in C^2\lxk$, or equivalently to $f_0$ admitting a continuously differentiable version on this region. Combining the preceding bounds yields the following guarantees.

\begin{restatable}[\bold{Piecewise-Linear Interpolation Error Bound}]{theorem}{CombinedErrorFspace}
\label{thm:combined_error_f}
Let $\Fhat$ be a cdf returned by Algorithm \ref{alg:main}. Suppose that $\Ttrue\in C^2\lxk$. Then, for every $\delta\in(0,1)$, with probability at least $1-\delta$,
\begin{align*}
\norm{\Fhat-\Ftrue}_{\lxk}\le R\left\{\sqrt{\frac{k\ln 3+\ln(2/\delta)}{\underline n}}+\sqrt{\frac{\ln(4k/\delta)}{2\underline n}}\right\}+\frac{R\Delta^2}{8}\norm{(1-\Ftrue)\Ttrue''}_{\lxk}.
\end{align*}
Moreover,
\begin{align*}
\norm{\Fhat-\Ftrue}_{[l,u]}\le R\left\{\sqrt{\frac{k\ln 3+\ln(2/\delta)}{\underline n}}+\sqrt{\frac{\ln(4k/\delta)}{2\underline n}}\right\}+\max\left\{\frac{R\Delta^2}{8}\norm{(1-\Ftrue)\Ttrue''}_{\lxk},1-\Ftrue(x_k)\right\}.
\end{align*}
\end{restatable}



The full-support bound separates three sources of error: stochastic estimation of the knot values, interpolation between the observed knots, and extrapolation beyond the final observed knot. Theorem~\ref{thm:combined_error_f} is therefore a refinement of Corollary~\ref{cor:baseline_error_f} that exploits the piecewise-linear reconstruction, rather than a uniformly stronger replacement: the factor $R$ can amplify the knotwise estimation error, but in exchange the interpolation term can decay at second order in $\Delta$. To quantify this improvement, suppose that along a sequence of designs with equal allocation and equidistant knot placement, the following additional conditions are satisfied\footnote{If these conditions are not satisfied, the bound remains valid but may exceed the trivial upper bound of one.}: $R=O(1)$, $\norm{(1-\Ftrue)\Ttrue''}_{\lxk}=O(1)$, and $1-\Ftrue(x_k)=O(\Delta^2)$. Then
\begin{align*}
\norm{\Fhat-\Ftrue}_{[l,u]}=O_p\left(\frac{k}{\sqrt N}+\frac{1}{k^2}\right).
\end{align*}

Thus, relative to the general bound in Corollary~\ref{cor:baseline_error_f}, the deterministic term improves from order $1/k$ to $1/k^2$, at the cost of the specified additional regularity conditions. These additional conditions, however, hold for a natural class of distributions. For example, suppose that $1-\Ftrue(x)=c(u-x)^\kappa$ for some $\kappa\ge2$ and $c=(u-l)^{-\kappa}$. Under an equidistant full-support grid,
\begin{align*}
R=2^\kappa,\qquad (1-\Ftrue(x))\Ttrue''(x)=-c\kappa(u-x)^{\kappa-2},\qquad 1-\Ftrue(x_k)=c\Delta^\kappa.
\end{align*}
Thus, $R$ and the survival-weighted curvature remain bounded as the grid is refined, while the terminal mass is $O(\Delta^2)$. Additionally, for any fixed $\alpha,\beta\ge2$, the Beta$(\alpha,\beta)$ distribution on $[0,1]$ also satisfies these conditions under equidistant full-support refinement.

By contrast, the uniform distribution on $[0,1]$ illustrates why the refined bound need not improve on the baseline guarantee. Its terminal survival mass is of order $\Delta$, while $(1-\Ftrue)\abs{\Ttrue''}$ diverges near the upper endpoint. The distribution nevertheless is covered by Corollary~\ref{cor:baseline_error_f}, which applies to all distributions. Thus, the baseline result remains useful when the additional regularity governing the piecewise-linear reconstruction is not available.

\subsection{Offline Data Collection Design}
\label{sec:data-collection}

Suppose that a practitioner has a total data collection budget of $N$ observations and must commit in advance to a sampling plan for learning the underlying distribution. Specifically, the practitioner must select the number of knots $k$, their locations $x_1,\ldots,x_k$, and the sample allocations $n_1,\ldots,n_k$. Data are then collected according to this fixed offline plan; we do not consider adaptive or online data collection. We use the general full-support guarantee in Corollary~\ref{cor:baseline_error_f} to derive robust design guidelines. These recommendations are obtained by optimizing or balancing terms in the proved upper bound and should therefore be interpreted as principled heuristics rather than claims of statistically optimal design:

\leavehalfline
\begin{enumerate}[(a)]
    \item Allocate samples equally across the knots.
    \item Use an equidistant grid over the full support.
    \item Let the number of knots be relatively small, even if the budget is large. 
\end{enumerate}
\leavehalfline

Guideline (a) follows directly from the stochastic term in Corollary~\ref{cor:baseline_error_f}, which depends on $\underline n=\min_{i=1,\ldots,k}n_i$. For fixed $N$ and $k$, equal allocation maximizes $\underline n$ and therefore minimizes the stochastic term in the bound.
The placement recommendation in guideline (b) requires more qualification. The exact deterministic term in Corollary~\ref{cor:baseline_error_f} is the largest between-knot cdf increment, $\max_{i=0,\ldots,k}\{\Ftrue(x_{i+1})-\Ftrue(x_i)\}$. If $\Ftrue$ were known, knot locations could instead be selected to make these increments approximately equal. In the absence of such information, an equidistant grid over the full support equalizes all $k+1$ intervals, and minimizes their largest width. For such a grid, $\Delta=(u-l)/(k+1)=u-x_k$. When $\Ftrue$ has a bounded density $f_0$, this gives
\begin{align*}
\max_{i=0,\ldots,k}\left\{\Ftrue(x_{i+1})-\Ftrue(x_i)\right\}\le\norm{f_0}_{[l,u]}\Delta.
\end{align*}
Thus, equidistant spacing provides a natural no-information default under the bounded-density upper bound, rather than a generally optimal placement rule. As a robustness check, the case studies in Section \ref{sec:case-study} examine practice-driven designs that depart from these guidelines. For example, the pricing study collects more observations at the regular price than at the sale and high prices, and places its knots within a plausible pricing range rather than uniformly across the full support. The estimator nevertheless performs well. 

Regarding guideline (c), under equal allocation, equidistant full-support knot placement, and bounded $f_0$, Corollary~\ref{cor:baseline_error_f} gives $\norm{\Fhat-\Ftrue}_{[l,u]}=O_p\left(k/\sqrt{N}+1/k\right)$. The first term increases with $k$ because each knot receives fewer observations, whereas the second decreases as the grid becomes finer. Balancing these two upper-bound terms suggests
\begin{align*}
k\asymp N^{1/4},\qquad n_i\asymp N^{3/4},\qquad \norm{\Fhat-\Ftrue}_{[l,u]}=O_p\left(N^{-1/4}\right),
\end{align*}
where $\asymp$ denotes the same asymptotic order as $N$ grows. Thus, even with a large budget, the baseline bound favors substantial sampling at each of a relatively small number of knots.
Under the stronger regularity conditions following Theorem~\ref{thm:combined_error_f}, balancing the terms in the refined upper bound instead suggests $k\asymp N^{1/6}$ and $n_i\asymp N^{5/6}$, yielding the faster guarantee $\norm{\Fhat-\Ftrue}_{[l,u]}=O_p\left(N^{-1/3}\right)$. We retain the baseline prescription as the robust default because it requires substantially weaker conditions on the ground-truth distribution. Across both regimes, the operational takeaway is the same: collect many observations at relatively few knots (e.g., prices, inspection ages, or dosage levels), consistent with practical constraints that limit the amount of experimentation possible.

\section{Additional Failure-Rate Properties}
\label{sec:extensions}
We now establish how our framework captures several other common failure-rate properties. In summary, our change-of-variables reformulation $t(x)=\ln(1-F(x))$, with suitable additional constraints, applies to (i) decreasing failure rates, (ii) failure rate average properties, (iii) generalized failure rates, and (iv) new better than used properties. 
Throughout this section, we primarily consider distributions supported on $[0,\infty)$, the standard domain for the properties below. In each subsection, we describe how to impose the property and reconstruct a corresponding cdf, omitting separate algorithm environments for brevity. Code implementations for all four properties are available in our GitHub repository.

\subsection{Decreasing Failure Rate}

Some systems exhibit a decreasing failure rate (DFR) early in their life cycles, as is often the case when failures reflect infant mortality.  Newly commissioned systems that survive an initial ``burn-in'' period may become less likely to fail \citep{proschan1963theoretical, holcomb1985infant}, and patients who survive the initial period after diagnosis may face a declining mortality risk \citep{clark2003survival}. Recall that $t'(x)=-h(x)$,  where $h(x) = f(x)/(1-F(x))$. Thus, DFR is equivalent to convexity of $t(x)$, so formulation \eqref{eqn:covprob} extends by replacing the concavity constraints \eqref{eq:covprob-ifr-constraint} with convexity constraints. Unlike in the IFR setting, if $y_k=n_k$ but some earlier knot has $y_i<n_i$, the DFR convexity constraints ensure that the finite-dimensional problem attains its optimum, so no separate optimization branch is needed. To reconstruct a full estimate, we interpolate piecewise linearly in $t$-space through $(0,0)$ and the fitted knot values, then append the linear tail $\hat t(x)=\hat\tau_k+s_k(x-x_k)$ for $x\geq x_k$, where $s_k:=(\hat\tau_k-\hat\tau_{k-1})/(x_k-x_{k-1})$ is the slope between the final two knots. When $s_k<0$, this continuation preserves monotonicity and convexity and ensures that $\hat t(x)\to-\infty$ as $x\to\infty$, yielding a continuous DFR cdf. If $s_k=0$, convexity and monotonicity force any continuation through the fitted values to remain flat beyond $x_k$, so those values cannot yield a proper DFR cdf.\footnote{The probability of this flat-slope case tends to zero as the sample size at each knot grows.} We can instead replace each $\hat\tau_i$ by $\hat\tau_i-\varepsilon x_i$ for $\varepsilon>0$ and repeat the interpolation and tail construction. The resulting continuous DFR cdf need not attain the optimal log-likelihood, but its log-likelihood approaches the optimum as $\varepsilon\to 0$.
Constant failure rates are handled similarly by requiring $t(x)$ to be affine with negative slope, corresponding to an exponential distribution.

\subsection{Failure Rate Average}

Let the cumulative failure rate up to time $x$ be $H(x) = \int_0^x h(z) dz$. The average failure rate is then $H(x)/x$. 
A distribution has an increasing failure rate average (IFRA) if $H(x)/x$ is non-decreasing in $x$ \citep{Birnbaum1966}. This condition is more general than IFR: if a distribution is IFR, then it is IFRA, but the converse need not hold. 
Intuitively, this property allows the instantaneous failure rate to decrease locally, provided that its average from the origin remains non-decreasing.

Since $t'(x) = -h(x)$ and $t(0)=0$, it follows that $H(x)=-t(x)$ and therefore $H(x)/x=-t(x)/x$.
Hence, a distribution is IFRA if and only if $-t(x)/x$ is non-decreasing for $x>0$. At the knots, this property can be imposed through the linear constraints
\begin{align*}
-\frac{t(x_i)}{x_i} \leq -\frac{t(x_{i+1})}{x_{i+1}},
\qquad i=1,\ldots,k-1,
\end{align*}
which replace constraints \eqref{eq:covprob-ifr-constraint} in formulation \eqref{eqn:covprob}.

These constraints are also preserved by piecewise-linear interpolation in $t$-space: if the ratios $-t(x_i)/x_i$ are non-decreasing at the knots, then $-t(x)/x$ is non-decreasing between each pair of adjacent knots under the piecewise-linear interpolant. When $y_k<n_k$, we can therefore construct a full distributional estimate by interpolating through $(0,0)$ and the optimized knot values $\vecThat$ and appending the linear tail $\That(x)=\tauhat_k-a(x-x_k)$ for $x\geq x_k$, where $a:=\max\{1/x_k,-\tauhat_k/x_k\}$. Any $a>0$ with $a\geq-\tauhat_k/x_k$ preserves IFRA at the join and throughout the tail, and the floor $1/x_k$ ensures $a>0$ when      $\tauhat_k=0$. When $y_k=n_k$ and an earlier knot has $y_j<n_j$, let $x_j$ be the last such knot. As in the IFR setting, we optimize through $x_j$ and append the logarithmic tail $\That(x)=\tauhat_j+a\ln\left((x_{j+1}-x)/(x_{j+1}-x_j)\right)$ for $x_j\leq x<x_{j+1}$,
where $a:=\max\{1,-\tauhat_j(x_{j+1}-x_j)/x_j\}$. Since $\That(x)\to-\infty$ as $x\to x_{j+1}$, setting $\Fhat(x)=1$ for $x\geq x_{j+1}$ yields a continuous IFRA cdf.

Finally, decreasing and constant failure rate averages can be handled by reversing the IFRA inequalities or imposing them as equalities, respectively. In the decreasing case, piecewise-linear interpolation can be followed by the tail $\hat t(x)=\hat\tau_k-a(x-x_k)$, where $0<a\leq-\hat\tau_k/x_k$ when $\hat\tau_k<0$. In the constant case, setting $\hat t(x)=-\lambda x$ for $\lambda>0$ yields an exponential distribution.

\subsection{Generalized Failure Rate}

\citet{lariviere2001selling} introduced the \emph{generalized failure rate} $g(x)$, defined for $x>0$ by $g(x):=xh(x)$.
A distribution has an increasing generalized failure rate (IGFR) if $g(x)$ is non-decreasing in $x$. This property is more general than IFR because every IFR distribution is IGFR \citep{lariviere2006note}, but the converse need not hold. IGFR is particularly useful in pricing because $g(x)$ represents the price elasticity of demand, so its monotonicity yields a single-crossing optimal-pricing condition even when the ordinary failure rate is not monotone. In standard second-price auctions with iid bidders, this same property makes expected revenue single-peaked in the reserve price and characterizes any interior optimal reserve by $g(r)=1$.

The IGFR property admits a convenient characterization after transforming both the cdf and the horizontal axis. Let $t(x):=\ln(1-F(x))$ as before, but in addition set $v:=\ln x$, and define $\phi(v):=t(e^v)$. Wherever the relevant derivatives exist,
\begin{align*}
\phi'(v)=e^v t'(e^v)=xt'(x)=-xh(x)=-g(x).
\end{align*}
Because $x\mapsto\ln x$ is increasing, $g(x)$ is non-decreasing in $x$ if and only if $\phi'(v)$ is non-increasing in $v$, or equivalently, if and only if $\phi$ is concave. Thus, just as IFR is equivalent to concavity of the log-survival function $t(x)$ when plotted against $x$, IGFR is equivalent to concavity of the same function when the horizontal axis is rescaled from $x$ to $\ln x$ \citep{lariviere2006note}.

Let $\tau_i:=t(x_i)$ for the positive observed knots $x_1<\cdots<x_k$. Concavity of $\phi$ can be imposed through the linear constraints
\begin{align*}
\frac{\tau_i-\tau_{i-1}}{\ln x_i-\ln x_{i-1}}
\geq
\frac{\tau_{i+1}-\tau_i}{\ln x_{i+1}-\ln x_i},
\qquad i=2,\ldots,k-1,
\end{align*}
which replace constraints \eqref{eq:covprob-ifr-constraint} in formulation \eqref{eqn:covprob}. Together with the usual monotonicity constraints on the knot values, these constraints preserve the convexity of the finite-dimensional likelihood-maximization problem.

After optimizing the relevant knot values, linearly interpolate between successive $(\ln x_i,\hat\tau_i)$ coordinates in log-$x$ space. This preserves IGFR between the fitted knots and leaves the fitted log-likelihood unchanged. We complete the reconstruction by specifying the lower and upper tails.
For the lower tail, if $\hat\tau_1=0$, set $\hat t(x)=0$ on $[0,x_1]$. Otherwise, set $\hat t(x)=\hat\tau_1(x/x_1)^a$ on $[0,x_1]$. If at least two knots are fitted and $\hat\tau_2<\hat\tau_1$, choose $a:=(\hat\tau_2-\hat\tau_1)/[\hat\tau_1(\ln x_2-\ln x_1)]>0$ to match the log-$x$ slopes at $x_1$; if only $x_1$ is fitted, take $a=1$.\footnote{If $\hat\tau_1<0$ and $\hat\tau_2=\hat\tau_1$, no continuous IGFR extension to the origin preserves both fitted values. The probability of this case tends to zero as the sample size at each knot grows. Perturbing the fitted values to $\hat\tau_i-\varepsilon x_i$ yields a continuous IGFR cdf whose log-likelihood approaches the optimum as $\varepsilon\to 0$.} For each fitted knot $x_i$, let $s_i$ denote the left slope of $\hat t(e^v)$ at $v=\ln x_i$, using the lower-endpoint segment when $i=1$.
For the upper tail, there are two cases. When $y_k<n_k$, append $\hat t(x)=\hat\tau_k+b\ln(x/x_k)$ for $x\geq x_k$, where $b:=\min\{-1,s_k\}<0$. This preserves IGFR and ensures that $\hat t(x)\to-\infty$ as $x\to\infty$. When $y_k=n_k$ and an earlier knot has $y_j<n_j$, let $x_j$ be the last such knot and optimize through $x_j$. Append $\hat t(x)=\hat\tau_j+c\ln\left((\ln x_{j+1}-\ln x)/(\ln x_{j+1}-\ln x_j)\right)$ for $x_j\leq x<x_{j+1}$, where $c:=\max\{1,-s_j(\ln x_{j+1}-\ln x_j)\}$. This preserves concavity in log-$x$ space and sends $\hat t(x)$ to $-\infty$ as $x\to x_{j+1}$. Finally, $\hat F(x)=1$ for $x\geq x_{j+1}$ yields a continuous IGFR cdf.

Finally, we consider decreasing generalized failure rate (DGFR) distributions, which are special cases of DFR distributions \citep{paul2005closure}. A nondegenerate continuous DGFR distribution must have a strictly positive lower support point $l$. When $l$ is known, we can impose DGFR by anchoring $t(l)=0$, reversing the IGFR log-space slope inequalities, and interpolating linearly in log-$x$ space. Continuing a negative final log-space slope gives a proper upper tail. If that slope is zero, replacing $\hat\tau_i$ by $\hat\tau_i-\varepsilon\ln(x_i/l)$ preserves log-space convexity and yields a continuous DGFR cdf with log-likelihood arbitrarily close to the optimum.
\subsection{New Better Than Used}

Finally, another relevant property compares, for $x',x''\geq 0$, the conditional survival probability $P(X \geq x' + x'' \mid X \geq x')$ with the unconditional survival probability $P(X \geq x'')$. The former measures the probability that a machine survives an additional $x''$ units of time, conditional on having already survived to time $x'$, while the latter measures the probability that a new machine survives at least $x''$ units of time. If $P(X \geq x' + x'' \mid X \geq x') \leq P(X \geq x'')$, then the distribution is said to have the ``new better than used'' (NBU) property \citep{barlow1996mathematical}. Equivalently, using the survival function, a distribution is NBU if $1-F(x'+x'') \leq (1-F(x')) (1-F(x''))$. NBU is more general than both IFRA and IFR: IFRA implies NBU, and IFR implies IFRA.

We can express this condition using the same change-of-variables approach. Since $e^{t(x)}=1-F(x)$, the NBU condition is equivalent to, for all $x',x''\geq 0$,
\begin{align*}
e^{t(x'+x'')} \leq e^{t(x')} e^{t(x'')}
\iff
t(x'+x'') \leq t(x') + t(x'').
\end{align*}
Thus, NBU is equivalent to subadditivity of $t(x)$. 

Suppose that the observed knots lie on a common equally spaced lattice anchored at zero.
This is a mild assumption and is trivially satisfied if all the knots are placed at rational numbers. Let $\mathcal G:=\{z_j=j\Delta:\: j=0,\ldots,m\}$ be such a lattice containing every observed knot, with $z_m=x_k$, and let $\tau_j:=t(z_j)$. Here, $\Delta$ denotes the lattice spacing, which may differ from the spacing between observed knots. We maximize the same binomial log-likelihood over these grid values, subject to the usual cdf monotonicity constraints and the new linear subadditivity constraints
\begin{align*}
\tau_{p+q}\leq\tau_p+\tau_q,\qquad p,q\in\{0,\ldots,m\},\quad p+q\leq m,
\end{align*}
which replace constraints \eqref{eq:covprob-ifr-constraint} in formulation \eqref{eqn:covprob}. The resulting problem remains a finite-dimensional convex optimization problem, although the number of NBU constraints grows quadratically in the grid size. Importantly, piecewise-linear interpolation preserves subadditivity on an equally spaced lattice.  To see this, write $x'=(p+\alpha)\Delta$ and $x''=(q+\beta)\Delta$, where $\alpha,\beta\in[0,1]$. If $\alpha+\beta\leq1$, the desired inequality follows by combining the discrete subadditivity inequalities at indices $p+q$ and $p+q+1$. If $\alpha+\beta>1$, it follows by combining the corresponding inequalities at indices $p+q+1$ and $p+q+2$. Consequently, the piecewise-linear interpolant through any feasible grid values is subadditive on every pair $x',x''\in[0,z_m]$ satisfying $x'+x''\leq z_m$.

It remains to extend the interpolant beyond the final grid point. When $y_k<n_k$, choose
\begin{align*}
s:=\min\left\{-\frac{1}{x_k},
\min_{\substack{0\leq p,q\leq m,\ p+q>m}}
\frac{\hat\tau_p+\hat\tau_q-\hat\tau_m}{(p+q-m)\Delta}\right\},
\end{align*}
and define $\hat t(x):=\hat\tau_m+s(x-z_m)$ for $x\geq z_m$. This tail preserves subadditivity and ensures that $\hat t(x)\to-\infty$ as $x\to\infty$. This yields a continuous NBU cdf.
When $y_k=n_k$ and an earlier knot has $y_j<n_j$, let $x_j$ be the last such knot. As in the IFR setting, we optimize through $x_j$ and use the preceding interpolation and linear-tail construction to obtain an initial subadditive function $\hat t_{\mathrm{init}}$. We set $\hat t(x)=\hat t_{\mathrm{init}}(x)$ for $x\leq x_{j+1}-\Delta$ and $\hat t(x)=\hat t_{\mathrm{init}}(x)+\ln((x_{j+1}-x)/\Delta)$ for $x_{j+1}-\Delta<x<x_{j+1}$. This preserves subadditivity and the fitted values through $x_j$, while ensuring that $\hat t(x)\to-\infty$ as $x\to x_{j+1}$. Setting $\hat F(x)=1$ for $x\geq x_{j+1}$ yields a continuous NBU cdf.  In both cases, the resulting estimate attains the supremum log-likelihood. 

Finally, the new-worse-than-used (NWU) property corresponds to superadditivity of $t(x)$ and can be imposed by reversing the NBU grid inequalities. Piecewise-linear interpolation preserves superadditivity. The grid constraints and monotonicity imply $\hat\tau_p+\hat\tau_q\leq\hat\tau_m$ whenever $p+q>m$, so $L:=\max_{0\leq p,q\leq m,\ p+q>m}(\hat\tau_p+\hat\tau_q-\hat\tau_m)/((p+q-m)\Delta)\leq0$. When $L<0$, a compatible linear upper tail can be appended with slope $s<0$ satisfying $s\geq L$. If $L=0$, replacing $\hat\tau_i$ by $\hat\tau_i-\varepsilon z_i$ preserves superadditivity and lowers the bound to $-\varepsilon$. Choosing $s=-\varepsilon$ then yields a continuous NWU cdf with log-likelihood arbitrarily close to the optimum.

\subsection{Visualization}

\begin{figure}[H]
\centering
\includegraphics[width=0.62\textwidth]{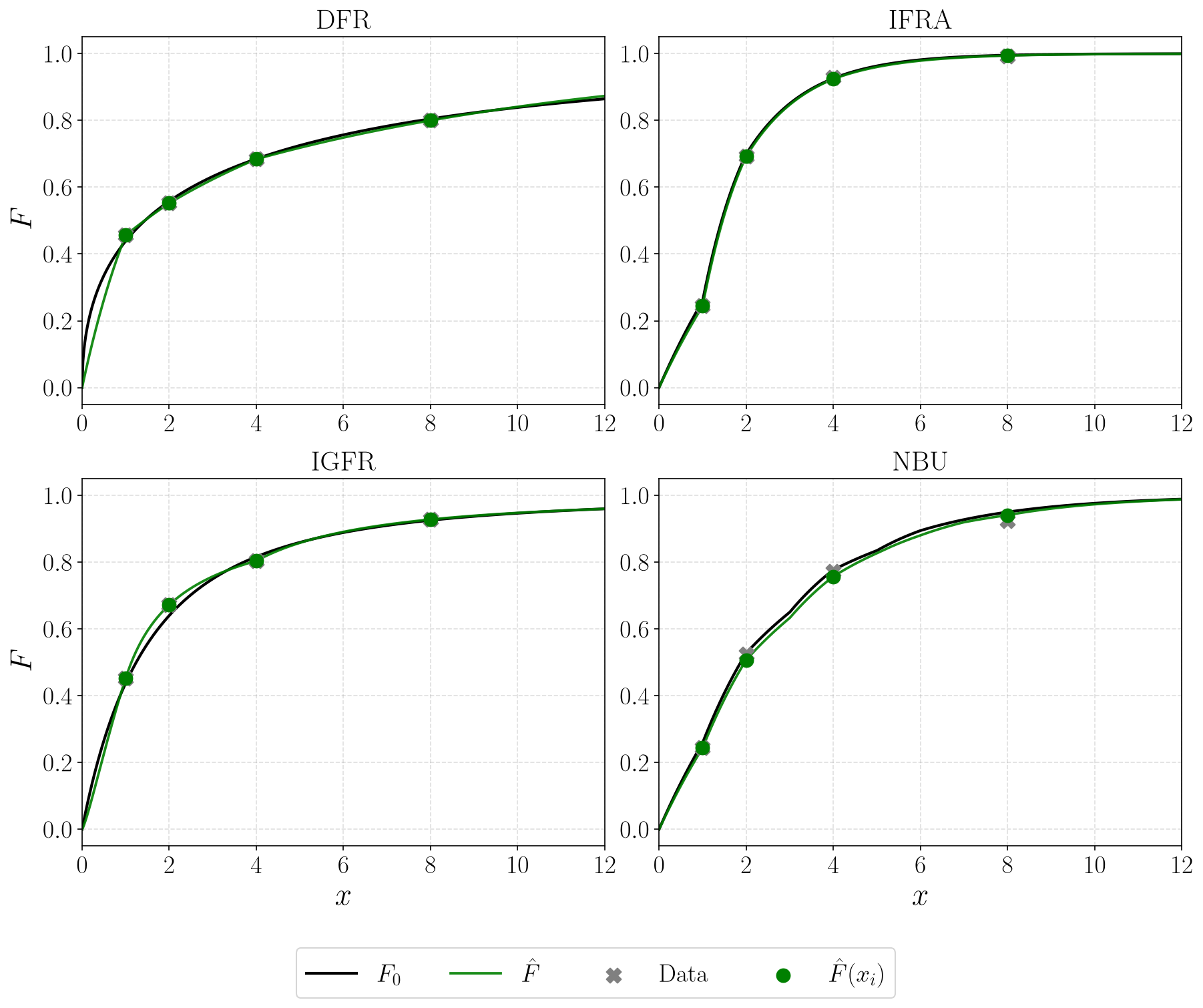}
\caption{\normalfont Numerical examples comparing $\Fhat$ against four ground-truth cdfs. For DFR, $\Ftrue$ is Weibull with shape $0.5$ and scale $3$. For IFRA, $\Ftrue$ has a piecewise-constant failure rate of $0.3$, $0.9$, and $0.7$ on $[0,1]$, $[1,2]$, and $[2,\infty)$, respectively. For IGFR, $\Ftrue$ is Lomax with shape $2$ and scale $3$. For NBU, $\Ftrue(x)=1-e^{-H(x)}$, where $H(x)=0.3\tilde H(x)$ and $\tilde H$ is piecewise linear through $(0,0)$, $(1,1)$, $(2,2.5)$, $(3,3.5)$, $(4,5)$, $(5,6)$, and $(6,7.5)$, with slope $1.25$ beyond $x=6$. Each example uses $k=4$ knots at $(1,2,4,8)$ with $n_i=250$ observations per knot.}
\label{fig:extensions}
\end{figure}

Figure~\ref{fig:extensions} illustrates the four estimators for the additional properties discussed above. In each panel, the ground-truth cdf satisfies the corresponding property but is not IFR. We use the same $k=4$ knots, $(1,2,4,8)$, and $n_i=250$ observations at each knot. Each estimate is obtained using the property-specific constraints, interpolation, and tail described above. Across all four examples, the estimated cdfs closely track the ground-truth cdfs.


\section{Numerical Experiments}
\label{sec:numerics}

In this section, we evaluate the numerical performance of Algorithm \ref{alg:main} for IFR distributions. Our experiments serve two purposes: (i) to validate the data-collection guidelines from Section \ref{sec:data-collection}, and (ii) to compare our estimator with two benchmarks in terms of estimation accuracy and computation time. We implement Algorithm \ref{alg:main} in Python and solve the convex program \eqref{eqn:covprob} using CVXPY \citep{diamond2016cvxpy, agrawal2018rewriting}. We construct the piecewise-linear interpolant of the estimated $\That$ symbolically using SymPy, and then apply the reverse mapping to obtain $\Fhat$.

Throughout this section, we consider $\text{Beta}(a,b)$ distributions with $a\ge 2$ and $b\ge 2$. These distributions are IFR \citep{bagnoli2005}, are naturally supported on $[0,1]$, and satisfy the conditions of Theorem \ref{thm:combined_error_f}. As a robustness check, we replicated all numerical experiments using truncated Normal, Gamma, and Weibull distributions and obtained qualitatively similar conclusions (details are omitted for brevity).

\subsection{Empirical Data Collection}
\label{sec:numerics-data-collection}

\begin{figure}[tbh]
\centering

\begin{subfigure}{0.31\textwidth}
    \centering
    \includegraphics[width=\linewidth]{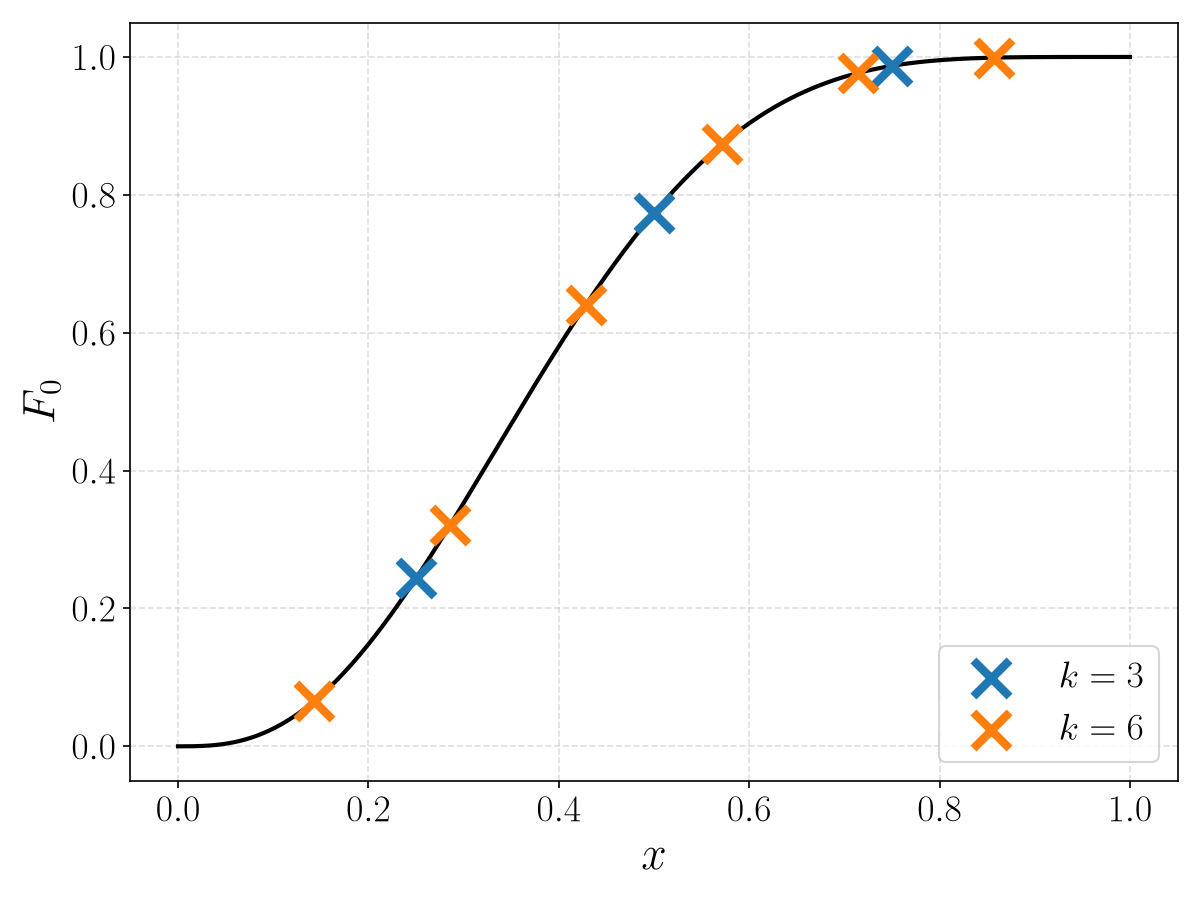}
    \caption{\footnotesize $F_0 \sim \text{Beta}(3,5)$}
\end{subfigure}
\hfill
\begin{subfigure}{0.31\textwidth}
    \centering
    \includegraphics[width=\linewidth]{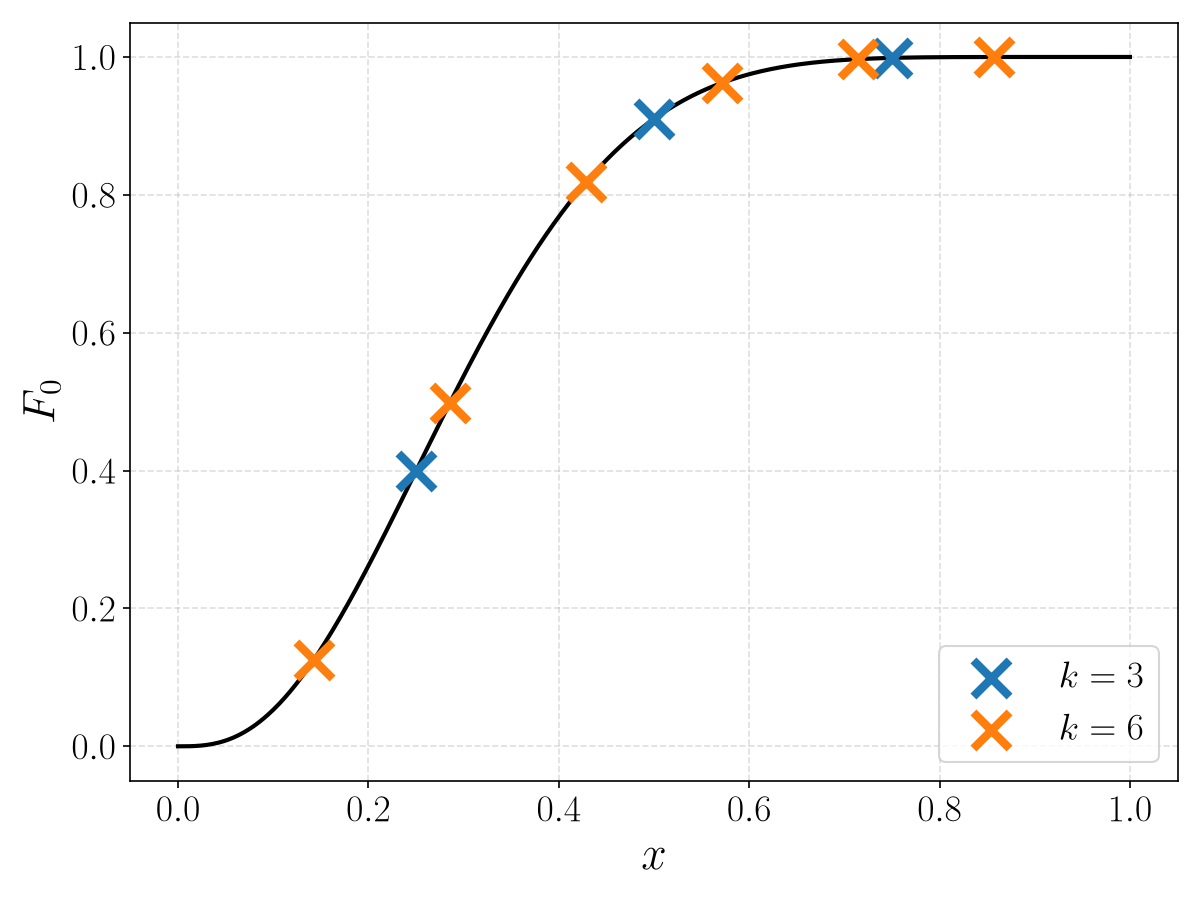}
    \caption{\footnotesize $F_0 \sim \text{Beta}(3,7)$}
\end{subfigure}
\hfill
\begin{subfigure}{0.31\textwidth}
    \centering
    \includegraphics[width=\linewidth]{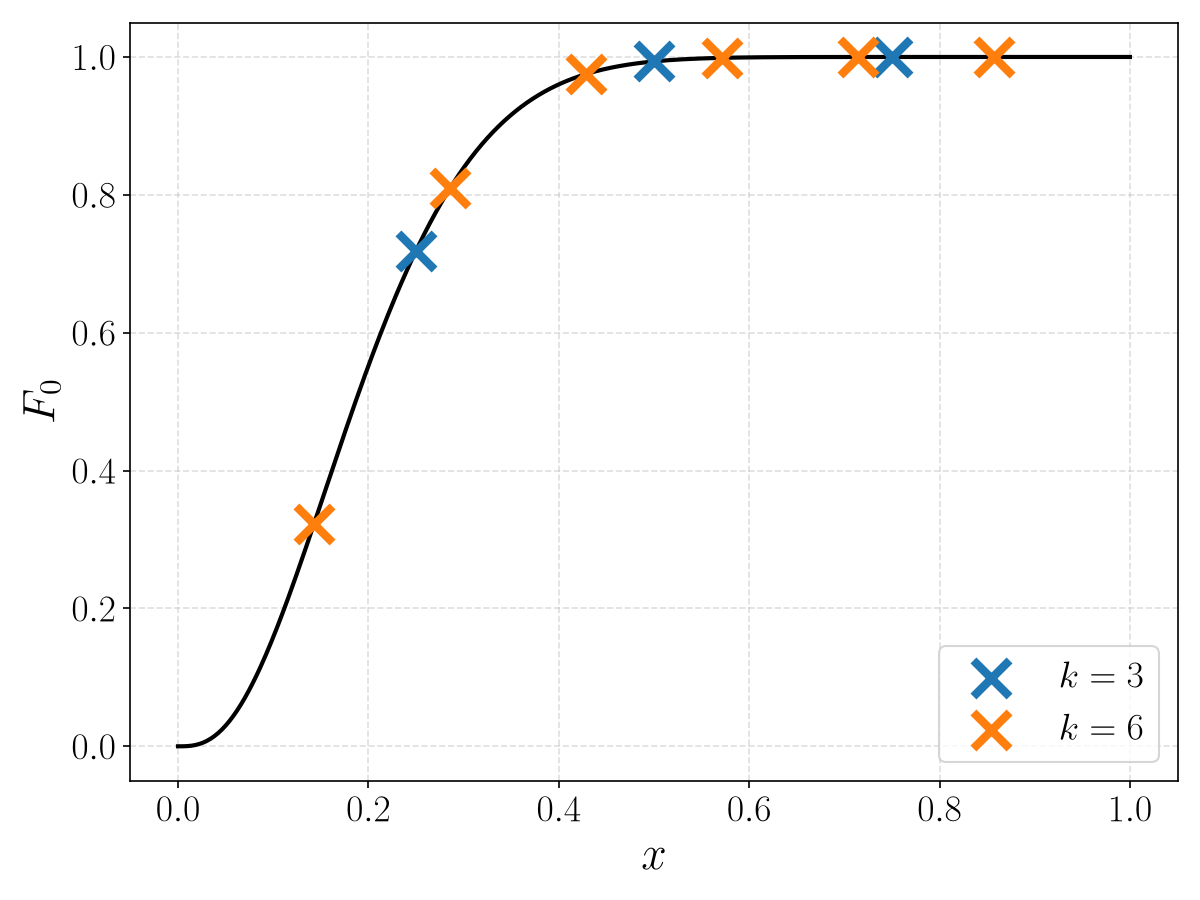}
    \caption{\footnotesize $F_0 \sim \text{Beta}(3,12)$}
\end{subfigure}

\vspace{0.75em}

\begin{subfigure}{0.31\textwidth}
    \centering
    \includegraphics[width=\linewidth]{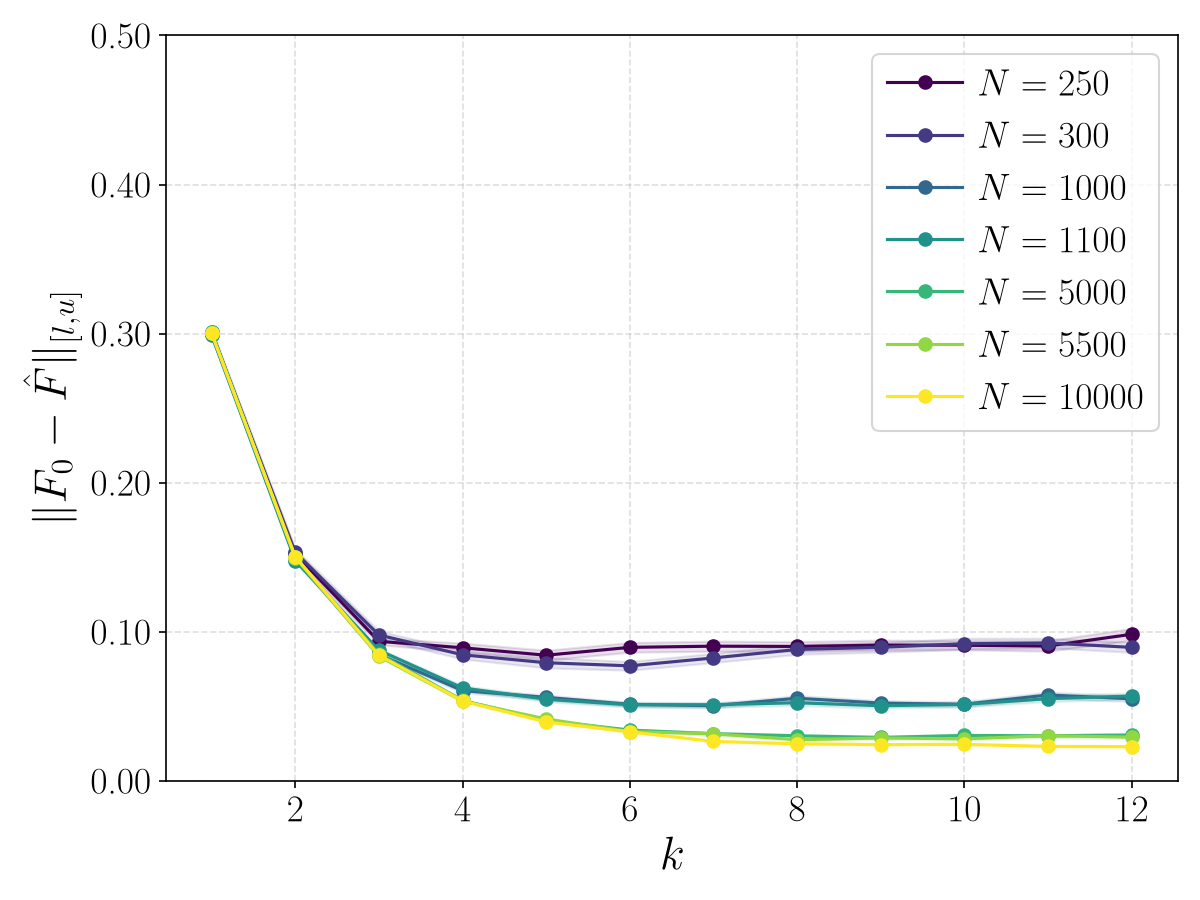}
    \caption{\footnotesize }
\end{subfigure}
\hfill
\begin{subfigure}{0.31\textwidth}
    \centering
    \includegraphics[width=\linewidth]{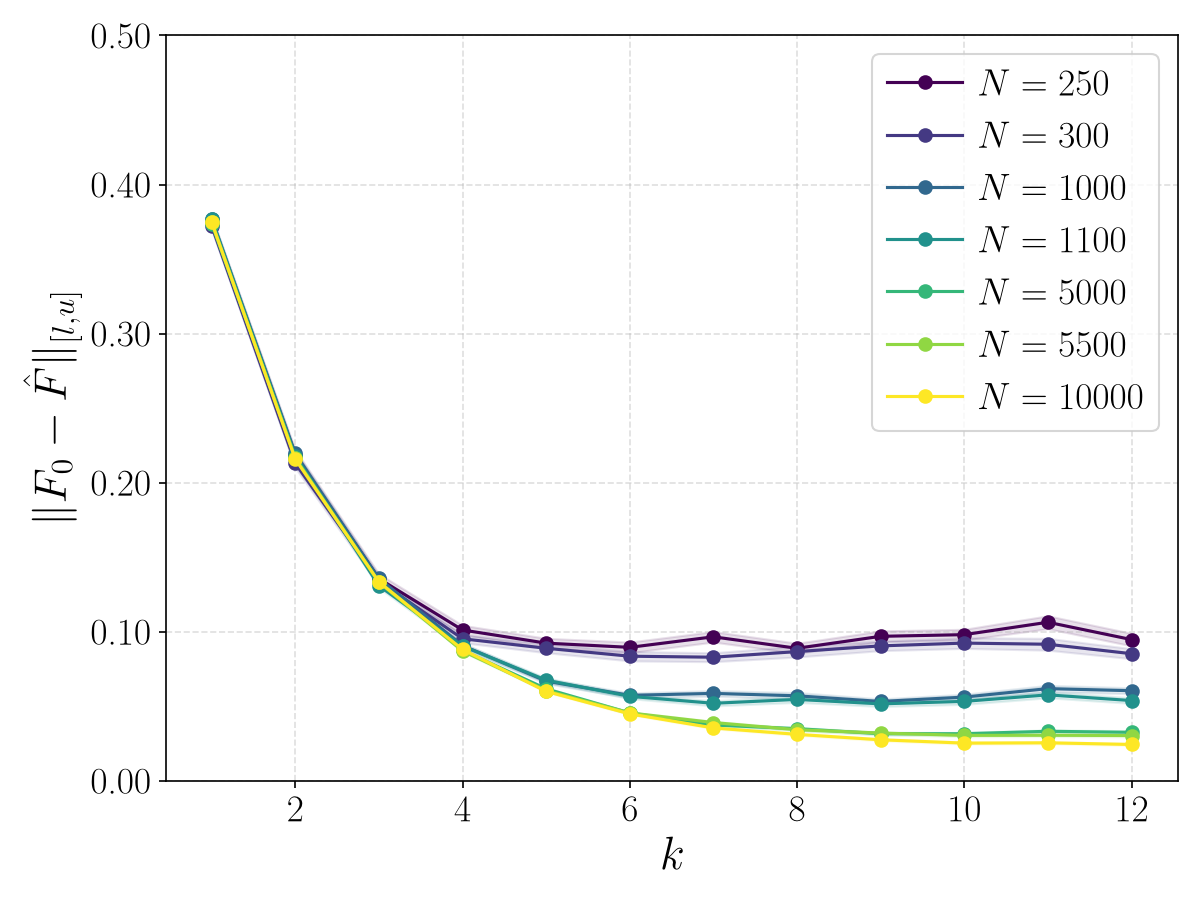}
    \caption{\footnotesize }
\end{subfigure}
\hfill
\begin{subfigure}{0.31\textwidth}
    \centering
    \includegraphics[width=\linewidth]{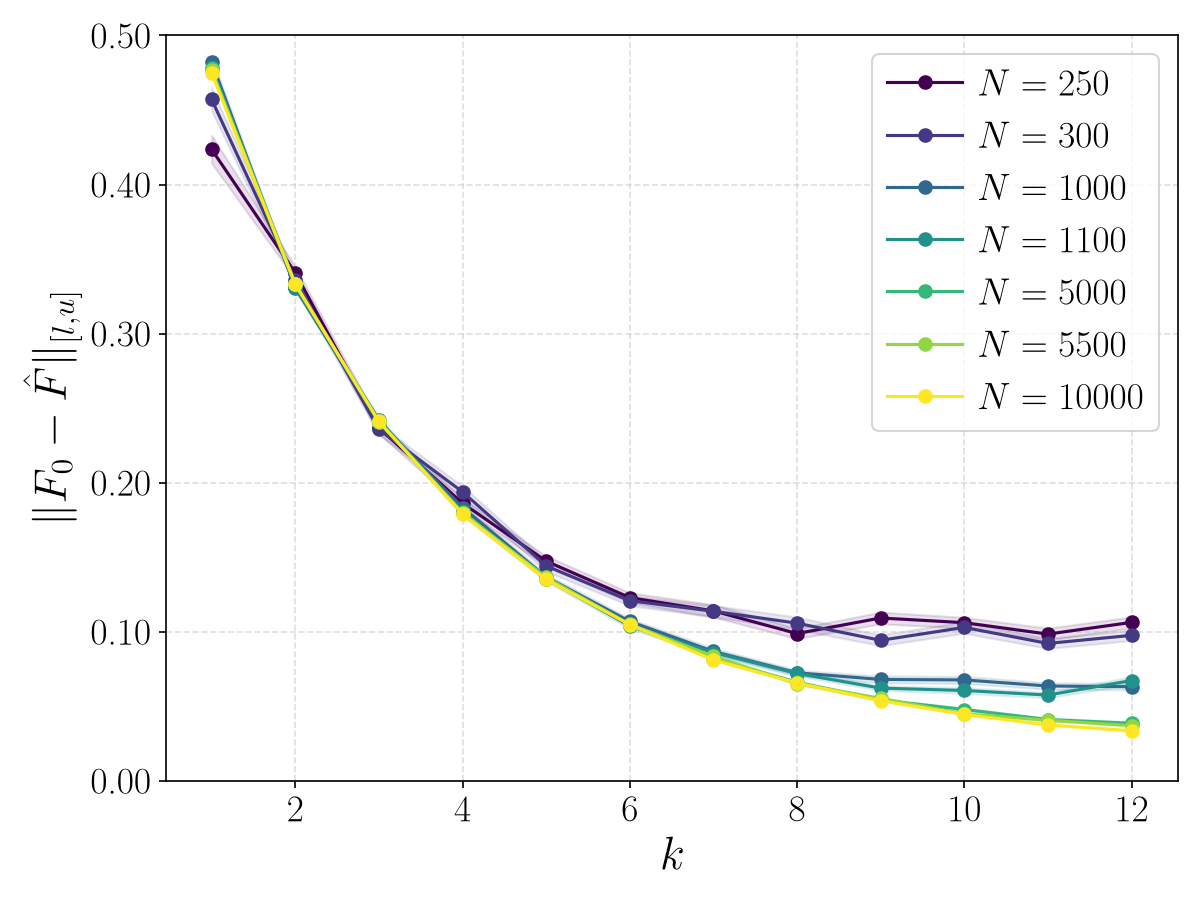}
    \caption{\footnotesize}
\end{subfigure}

\caption{\normalfont Comparison of $\Fhat$ from Algorithm \ref{alg:main} with ground-truth cdfs $F_0\sim\text{Beta}(3,b)$ for $b\in\{5,7,12\}$ under fixed total sample budgets. (Top) Cdfs with the locations of 3 and 6 equidistant knots; (Bottom) $\norm{\Ftrue-\Fhat}_{[l,u]}$ as the number of knots varies. Shaded regions represent standard errors over 100 runs.}
\label{fig:constant-N}
\end{figure}

We examine how the number of knots affects estimation accuracy under a fixed total sampling budget $N$. For each design, we place $k$ equidistant interior knots, and each knot receives $\lfloor N/k\rfloor$ observations. We consider low, medium, high, and very high data regimes, represented by increasing values of $N$. Figure \ref{fig:constant-N} considers three ground-truth cdfs, $\text{Beta}(3,b)$ with $b\in\{5,7,12\}$. As $b$ increases, the distribution becomes more concentrated near the lower endpoint and its cdf rises more sharply over a narrower region. Panels (a)--(c) depict the three cdfs and illustrative knot locations. Panels (d)--(f) plot the error $\norm{\Ftrue-\Fhat}_{[l,u]}$ as $k$ varies under each fixed total sample budget.

Across all three distributions, the error decreases sharply as $k$ initially increases and then enters a nearly flat region once the knot set adequately captures the shape of the underlying cdf. For example, under the smaller budgets, the $\mathrm{Beta}(3,5)$ cdf reaches this region at approximately $k=3$, whereas the more concentrated $\mathrm{Beta}(3,12)$ cdf requires approximately $k=6$. As $N$ increases, the number of knots associated with this near-plateau increases modestly, but the change is small relative to the increase in the sampling budget. This pattern reinforces the third data-collection guideline in Section \ref{sec:data-collection}: as the sampling budget increases, it should be used primarily to increase the number of observations at existing knots rather than to continually refine the grid. Increasing $N$ also reduces error at every value of $k$, although the incremental reductions diminish in the larger-budget regimes, as reflected by the increasingly small gaps between the corresponding curves. Finally, increasing $k$ beyond the near-plateau does not substantially degrade performance despite spreading a fixed budget across more knots. This suggests that the additional knots help the estimator explore the cdf's shape, while the IFR constraint provides enough structure to maintain stable performance. Overall, although the precise choice of $k$ depends on the underlying distribution, the numerics support the broader guideline that relatively few knots are sufficient under a fixed sampling budget.

\vspace{-1cm}
\subsection{Benchmark Comparisons}
\label{sec:numerics-benchmarks}

\begin{figure}[tbh]
    \centering
    \includegraphics[width=0.55\linewidth]{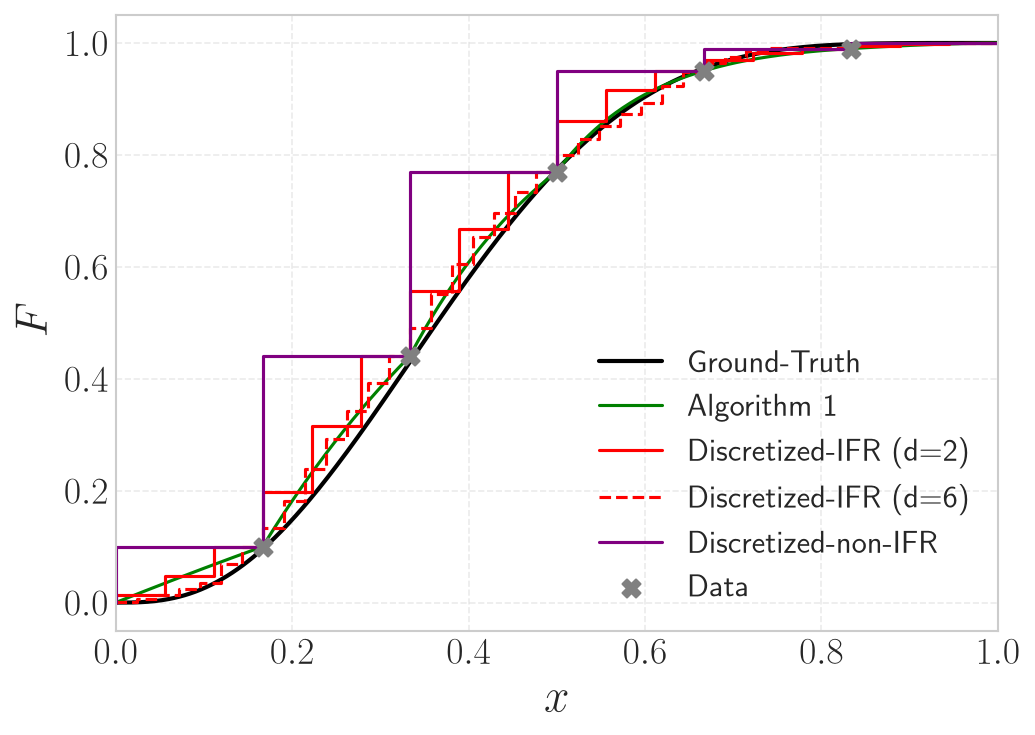}
    \caption{\normalfont Illustration of the fitted cdfs from Algorithm \ref{alg:main}, discretized-IFR with $d\in\{2,6\}$, and discretized-non-IFR. The ground-truth cdf is $F_0\sim\text{Beta}(3,5)$, with $k=5$ knots and $n_i=100$ observations per knot.}
    \label{fig:visualization-benchmarks}
\end{figure}

Next, we compare the estimation accuracy and computation time of Algorithm \ref{alg:main} with two benchmarks. The first, discretized-IFR, solves a discretized version of the original IFR-constrained problem \eqref{eqn:originalprob}. The second, discretized-non-IFR, solves the same discretized problem without the IFR constraint \eqref{eq:origprob-ifr-constraint}, yielding a monotone cdf analogous to a binomial version of isotonic regression \citep{ayer1955empirical}. For both benchmarks, we construct a piecewise-constant step function\footnote{Step-function cdf estimators are standard in nonparametric statistics. In the discretized-IFR benchmark, a sufficiently dense grid with larger $d$ makes the step-function reconstruction a close approximation to a continuous reconstruction, such as linear interpolation in $F$-space.} to recover a full cdf estimate. The discretized-IFR benchmark assesses the computational value of the $t=\ln(1-F)$ change-of-variables reformulation, whereas the discretized-non-IFR benchmark assesses the value of imposing IFR, rather than monotonicity alone. We solve both benchmark problems using Gurobi \citep{gurobi}. For discretized-IFR, we express the IFR constraint as a bilinear constraint, equivalently a nonconvex quadratic constraint, as discussed at the end of Section \ref{sec:problem}. To manage its computational burden, we enforce the constraint on a reduced grid. Let $d$ denote the number of auxiliary grid points inserted between each pair of adjacent observed knots. When $d=0$, the IFR constraints are imposed only at the observed knots; increasing $d$ refines the grid and adds variables and constraints.\footnote{To help the optimizer, we set an upper bound on the objective equal to the unconstrained binomial log-likelihood, namely $\sum_{i=1}^k \left(y_i \ln(y_i/n_i) + (n_i-y_i)\ln(1-y_i/n_i)\right)$.} Throughout this comparison, the ground-truth cdf is $F_0\sim\text{Beta}(3,5)$. Figure \ref{fig:visualization-benchmarks} illustrates the fitted cdfs from Algorithm \ref{alg:main}, discretized-IFR, and discretized-non-IFR for one realization of the noisy quantile data.

\begin{figure}[tbh]
    \centering
    \includegraphics[width=0.9\linewidth]{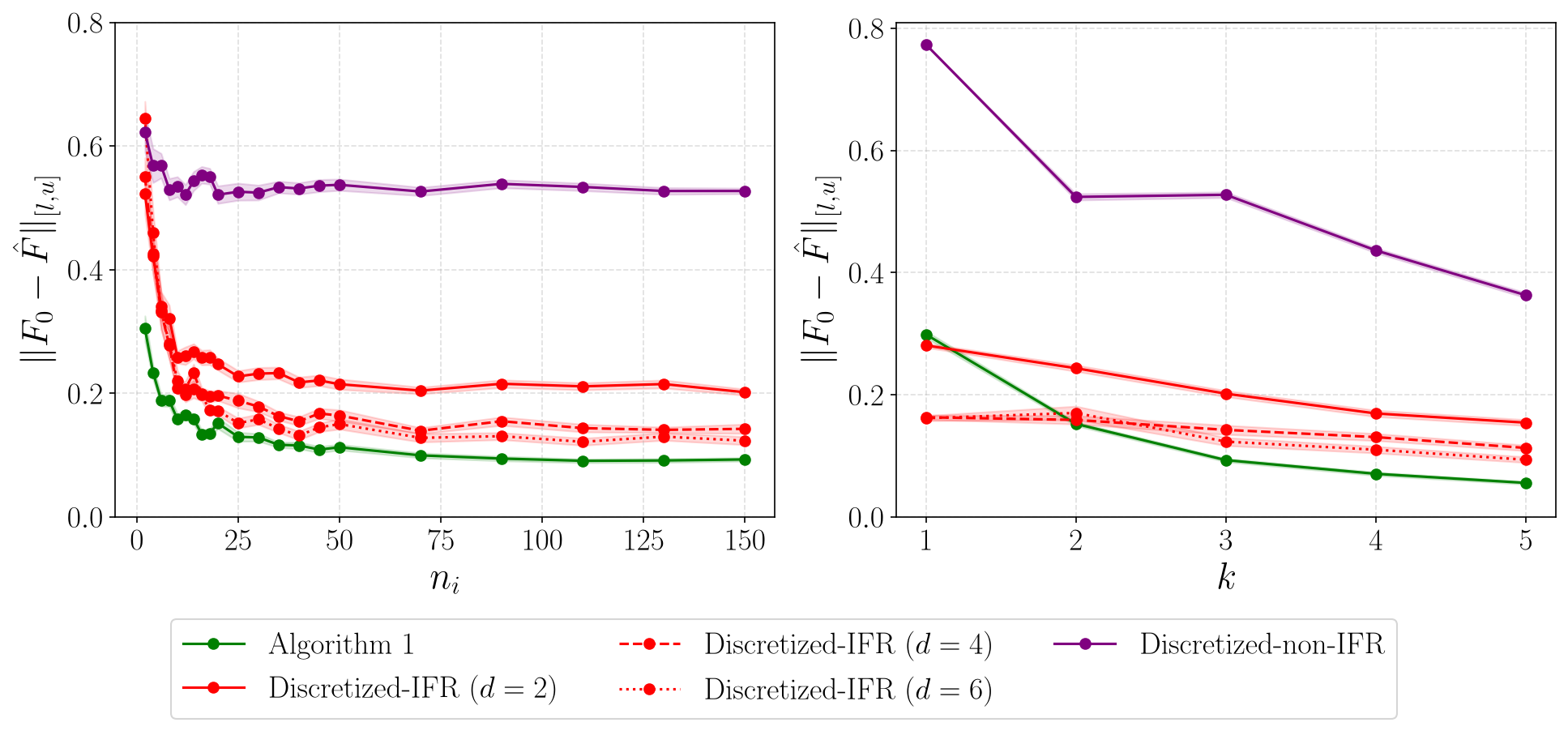}
    \caption{\normalfont Error $\norm{\Ftrue-\Fhat}_{[l,u]}$ for Algorithm \ref{alg:main}, discretized-IFR with $d\in\{2,4,6\}$, and discretized-non-IFR under $F_0\sim\text{Beta}(3,5)$. (Left) fixes $k=3$ and varies $n_i$; (Right) fixes $n_i=150$ and varies $k$. Shaded regions represent standard errors over 50 runs.}
    \label{fig:method-comparison}
\end{figure}

We first compare estimation accuracy across repeated instances. We fix $k=3$ and vary $n_i$ from 2 to 150, and then fix $n_i=150$ and vary $k$ from 1 to 5. Figure \ref{fig:method-comparison} reports the resulting errors $\norm{\Ftrue-\Fhat}_{[l,u]}$ over 50 runs. As the number of observations per knot increases, the error decreases substantially for Algorithm \ref{alg:main} and for each discretized-IFR variant. Increasing the number of knots also improves accuracy, although the gains are more modest over the plotted range. Algorithm \ref{alg:main} consistently improves upon discretized-non-IFR and generally outperforms the coarser discretized-IFR variants. Increasing $d$ typically improves the accuracy of discretized-IFR, and the finest discretization is competitive with, and occasionally improves upon, Algorithm \ref{alg:main}. These statistical gains, however, come at a substantially greater computational cost, which we discuss next.

For each instance, we impose a maximum solve time of 300 seconds. Table \ref{tab:solve_times} reports mean solution times and the number of instances that fail to solve within this limit.\footnote{For discretized-IFR, if a run did not return an accepted solution at density $d$, we did not attempt that run at larger densities $d'>d$, because we observed that such instances typically also failed at those densities. Consequently, we assign a computation time of 300 seconds to both failed attempts and unattempted instances with the same data at larger densities. When computing error performance, however, we consider only successfully optimized instances, which biases the reported performance in favor of this benchmark.} We assign each failed instance a solve time of 300 seconds when computing the reported mean. Therefore, the reported means are conservative lower bounds on the true mean solution times. Algorithm \ref{alg:main} and discretized-non-IFR solve in a fraction of a second throughout the experiment. In contrast, discretized-IFR is fast only at low grid densities. Its computation time grows rapidly with $d$, and the finer discretizations fail to solve within the time limit in several instances. Thus, although grid refinement can improve the statistical accuracy of discretized-IFR, the corresponding nonconvex optimization problem quickly becomes computationally impractical.

\begin{table}[H]
\centering
\small
\begin{tabular}{cc|c|ccc|c}
\toprule
 & & \multicolumn{1}{c|}{\bold{Algorithm 1}} 
 & \multicolumn{3}{c|}{\bold{Discretized-IFR}} 
 & \multicolumn{1}{c}{\bold{Discretized-}} \\
\bold{$k$} & \bold{$n_i$} 
& 
& \bold{$d=2$} 
& \bold{$d=4$} 
& \bold{$d=6$} 
& \bold{non-IFR} \\
\midrule
 & 2 & 0.01 $\pm$ 0.01 & 0.00 $\pm$ 0.00 & 0.00 $\pm$ 0.00 & 0.00 $\pm$ 0.00 & 0.00 $\pm$ 0.00 \\
 & 10 & 0.02 $\pm$ 0.00 & 0.01 $\pm$ 0.00 & 0.01 $\pm$ 0.00 & 0.01 $\pm$ 0.00 & 0.00 $\pm$ 0.00 \\
1 & 20 & 0.02 $\pm$ 0.00 & 0.01 $\pm$ 0.00 & 0.01 $\pm$ 0.00 & 0.01 $\pm$ 0.00 & 0.00 $\pm$ 0.00 \\
 & 50 & 0.02 $\pm$ 0.00 & 0.01 $\pm$ 0.00 & 0.01 $\pm$ 0.00 & 0.01 $\pm$ 0.00 & 0.00 $\pm$ 0.00 \\
\midrule
 & 2 & 0.02 $\pm$ 0.01 & 0.06 $\pm$ 0.06 & 0.31 $\pm$ 0.51 & 3.49 $\pm$ 9.44 & 0.01 $\pm$ 0.00 \\
 & 10 & 0.02 $\pm$ 0.01 & 0.13 $\pm$ 0.13 & 0.42 $\pm$ 1.02 & 1.05 $\pm$ 3.04 & 0.00 $\pm$ 0.00 \\
3 & 20 & 0.03 $\pm$ 0.01 & 0.09 $\pm$ 0.09 & 0.44 $\pm$ 1.19 & 2.40 $\pm$ 12.07 & 0.01 $\pm$ 0.00 \\
 & 50 & 0.02 $\pm$ 0.00 & 0.05 $\pm$ 0.04 & 0.11 $\pm$ 0.08 & 0.27 $\pm$ 0.35 & 0.00 $\pm$ 0.00 \\
\midrule
 & 2 & 0.02 $\pm$ 0.01 & 2.45 $\pm$ 9.58 & 20.55 $\pm$ 51.17 & 47.98 $\pm$ 92.40 (4) & 0.01 $\pm$ 0.01 \\
 & 10 & 0.02 $\pm$ 0.00 & 2.58 $\pm$ 9.91 & 14.27 $\pm$ 46.87 (1) & 50.54 $\pm$ 83.13 (3) & 0.01 $\pm$ 0.02 \\
5 & 20 & 0.02 $\pm$ 0.00 & 1.92 $\pm$ 3.35 & 22.32 $\pm$ 51.13 (1) & 99.44 $\pm$ 123.02 (10) & 0.01 $\pm$ 0.02 \\
 & 50 & 0.03 $\pm$ 0.01 & 0.83 $\pm$ 1.65 & 10.25 $\pm$ 21.36 & 62.88 $\pm$ 99.91 (4) & 0.01 $\pm$ 0.00 \\
\bottomrule
\end{tabular}
\leavequarterline
\caption{\normalfont Mean solution times, in seconds, for the estimation methods under the ground-truth cdf $F_0\sim\text{Beta}(3,5)$. Unsolved instances are assigned a solution time of 300 seconds. Cells report the mean $\pm$ standard deviation; parenthetical values, shown only when positive, report the number of unsolved instances.}
\label{tab:solve_times}
\end{table}


\section{Case Studies}
\label{sec:case-study}

This section presents two synthetic, application-motivated case studies in revenue management and reliability. Each case study uses a fixed, domain-specific design for knot locations and sample allocation, and evaluates an estimate-then-optimize policy over repeated simulated data sets. Our objectives are to assess the estimation accuracy of Algorithm \ref{alg:main} and to evaluate the quality of its resulting downstream decisions under the ground-truth cdf. We focus on settings in which a fully IFR reconstruction preserves a useful structural property of the estimated objective, namely unimodality of the revenue or replacement-cost function.

\subsection{Monopolistic Pricing}
Consider a firm selling an indivisible good to customers whose willingness-to-pay values are iid draws from an unknown IFR cdf $F_0$ supported on $[20,50]$. The firm collects sale or no-sale observations at three prices: a sale price of $\$25$ with 50 observations, a regular price of $\$30$ with 100 observations, and a high price of $\$35$ with 50 observations. The firm seeks to learn the revenue-maximizing posted price \citep{Gallego2019}. If $F_0$ were known, expected revenue at price $p$ would be $\mathcal{R}(p)=p(1-F_0(p))$. The IFR property implies that $\mathcal{R}$ is unimodal \citep{lariviere2006note}. Let the optimal price be denoted by
\begin{align*}
p^* \in \argmax_{p\in[20,50]} p(1-F_0(p)).
\end{align*}

In practice, the firm follows an estimate-then-optimize strategy. It first constructs an estimate $\Fhat$ and then selects a price $\hat{p}\in\argmax_{p\in[20,50]}p(1-\Fhat(p))$. We compare Algorithm \ref{alg:main} with discretized-IFR using $d=2$ and discretized-non-IFR. We compute both $p^*$ and $\hat{p}$ over a fine grid of candidate prices.\footnote{When the candidate grid contains multiple maximizing prices, we select the smallest maximizer.} Writing $\mathcal R(p;F):=p(1-F(p))$, we evaluate the expected performance of $\hat p$ under the ground-truth cdf using the revenue ratio
\begin{align*}
\frac{\mathcal R(\hat p;\Ftrue)}{\mathcal R(p^*;\Ftrue)} = 
\frac{\hat{p}(1-F_0(\hat{p}))}{p^*(1-F_0(p^*))},
\end{align*}
which lies in $[0,1]$, with larger values indicating better performance. We next relate the full-support cdf error studied in Section \ref{sec:theory} to this ratio (see \cite{besbes2025beyond} for a similar bound).
\begin{restatable}[\bold{Revenue Ratio Bound}]{proposition}{PricingRevenueRatio}
Let $\varepsilon:=\norm{\Fhat-\Ftrue}_{[l,u]}$. Then, the revenue ratio satisfies
\begin{align*}
\frac{\mathcal R(\hat p;\Ftrue)}{\mathcal R(p^*;\Ftrue)} \ge \left[1-\frac{(p^*+\hat p)\varepsilon}{\mathcal R(p^*;\Ftrue)}\right]_+ \geq \left[1-\frac{2u}{l}\varepsilon\right]_+.
\end{align*}
\label{prop:revenue-ratio}
\end{restatable}

\vspace{-0.8cm}
We simulate 150 runs with four ground-truth cdfs: truncated Chi-square, Normal, Gamma, and Logistic distributions, all supported on $[20,50]$. With $l=20$ and $u=50$, Proposition \ref{prop:revenue-ratio} guarantees a revenue ratio of at least $\left[1-5\varepsilon\right]_+$. For example, if $\varepsilon = 0.05$, the revenue ratio bound is 0.75.


\begin{figure}[tbh]
\centering
\includegraphics[width=0.88\linewidth]{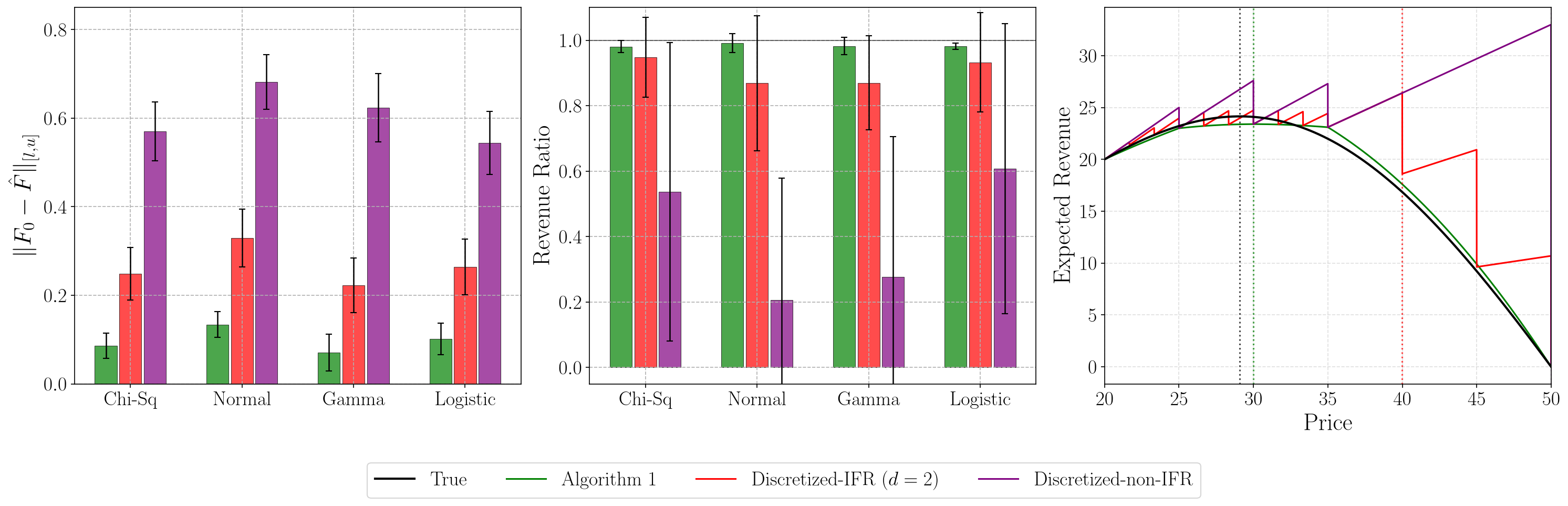}
\caption{\normalfont Pricing case study over 150 runs with four ground-truth truncated cdfs supported on $[20,50]$: $\Ftrue\sim\text{Chi-Square}(38)$, $\Ftrue\sim\text{Normal}(38,6)$, $\Ftrue\sim\text{Gamma}(8,6)$, and $\Ftrue\sim\text{Logistic}(36,5)$. (Left) Error $\norm{\Ftrue-\Fhat}_{[l,u]}$. (Middle) Corresponding revenue ratios, with the horizontal line indicating the oracle ratio of one. The bars show means and error bars show one sample standard deviation. (Right) Representative revenue curves for a Gamma instance, with dotted vertical lines denoting the selected prices.}
\label{fig:pricing_case_study}
\end{figure}

Figure \ref{fig:pricing_case_study} reports the error $\norm{\Ftrue-\Fhat}_{[l,u]}$ and the corresponding revenue ratio across our simulations. Algorithm \ref{alg:main} achieves the lowest mean full-support error across all four ground-truth cdfs. The discretized-IFR benchmark has intermediate error, whereas discretized-non-IFR produces substantially larger errors and greater variability. Moreover, Algorithm \ref{alg:main} yields revenue ratios close to one across all four distributions, with limited run-to-run variation. Discretized-IFR delivers lower average revenue ratios and greater variability, while discretized-non-IFR can prescribe prices that lead to substantial revenue loss. Finally, the right panel in Figure \ref{fig:pricing_case_study} illustrates the mechanism in a representative Gamma instance. Algorithm \ref{alg:main} induces a unimodal revenue curve with a selected price close to the true optimum. The discretized benchmarks can induce multimodal revenue curves because their reconstructions do not guarantee a fully IFR cdf between grid points.



\subsection{Preventive Maintenance}


Consider a firm responsible for a fleet of remotely deployed automated external defibrillators (AEDs). Each AED has an iid time to failure with unknown IFR cdf $F_0$ supported on $[0,12]$, measured in months. The AEDs are distributed across many locations, each of which follows one of three inspection schedules. Each AED is inspected exactly once at age 2, 4, or 6 months, and the inspection reveals only whether it has failed or remains operational at that age. We observe 50 AEDs at each inspection age.

The firm wishes to implement an age-replacement policy under which each AED is preventively replaced at age $\kappa$, unless it fails earlier, in which case it is replaced immediately \citep{jardine2013maintenance}. Let $c_p=300$ and $c_f=1200$ denote the costs of preventive and failure-based replacement, respectively. For a given cdf $F$, the long-run expected replacement cost per unit time under replacement age $\kappa$ is
\begin{subequations}
\begin{align}
    C(\kappa; F) 
    &= \frac{\text{total expected replacement cost per cycle}}{\text{expected cycle length}} \label{eq:PM-a}\\
    &= \frac{c_p(1-F(\kappa)) + c_fF(\kappa)}{\kappa(1-F(\kappa)) + \int_0^\kappa t f(t) dt} \label{eq:PM-b}\\
    &= \frac{c_p(1-F(\kappa)) + c_fF(\kappa)}{\int_0^\kappa (1-F(t)) dt}. \label{eq:PM-c}
\end{align}
\end{subequations}
Equation \eqref{eq:PM-a} follows from the renewal-reward formulation of age-replacement models \citep{smith1955}. Equation \eqref{eq:PM-b} separates cycles that end in preventive replacement from those that end in failure. Equation \eqref{eq:PM-c} follows by integration by parts, allowing the objective to depend only on the cdf and model primitives. The IFR property implies that $C(\kappa;F_0)$ is unimodal \citep{barlow1996mathematical}. Let an optimal replacement age be denoted by
\begin{align*}
\kappa^* \in \argmin_{\kappa\in(0,12]} C(\kappa;F_0).
\end{align*}

In practice, the firm first constructs an estimate $\hat{F}$ and then selects $\hat{\kappa}\in\argmin_{\kappa\in(0,12]} C(\kappa;\hat{F})$. We compute both $\kappa^*$ and $\hat{\kappa}$ over a fine grid of candidate replacement ages.\footnote{When the candidate grid contains multiple minimizing replacement ages, we select the smallest minimizer.} We evaluate the expected performance of $\hat\kappa$ under the ground-truth cdf using the cost ratio
\begin{align*}
\frac{C(\kappa^*;\Ftrue)}{C(\hat\kappa;\Ftrue)},
\end{align*}
which lies in $[0,1]$, with larger values indicating better performance. We next relate the full-support cdf error studied in Section \ref{sec:theory} to this ratio.
\begin{restatable}[\bold{Preventive Maintenance Cost Ratio}]{proposition}{PreventiveMaintenanceCostRatio}
Let $\varepsilon:=\norm{\Fhat-\Ftrue}_{[l,u]}$, $\mu_0:=\int_0^u(1-\Ftrue(t))\,dt$, $a:=(c_f-c_p)\varepsilon/c_p$, and $b:=u\varepsilon/\mu_0$. If $a<1$ and $b<1$, then the cost ratio satisfies
\begin{align*}
\frac{C(\kappa^*;\Ftrue)}{C(\hat\kappa;\Ftrue)}
\geq \frac{(1-a)(1-b)}{(1+a)(1+b)}.
\end{align*}
\label{prop:cost-ratio}
\end{restatable}
%

\begin{figure}[tbh]
\centering
\includegraphics[width=0.88\linewidth]{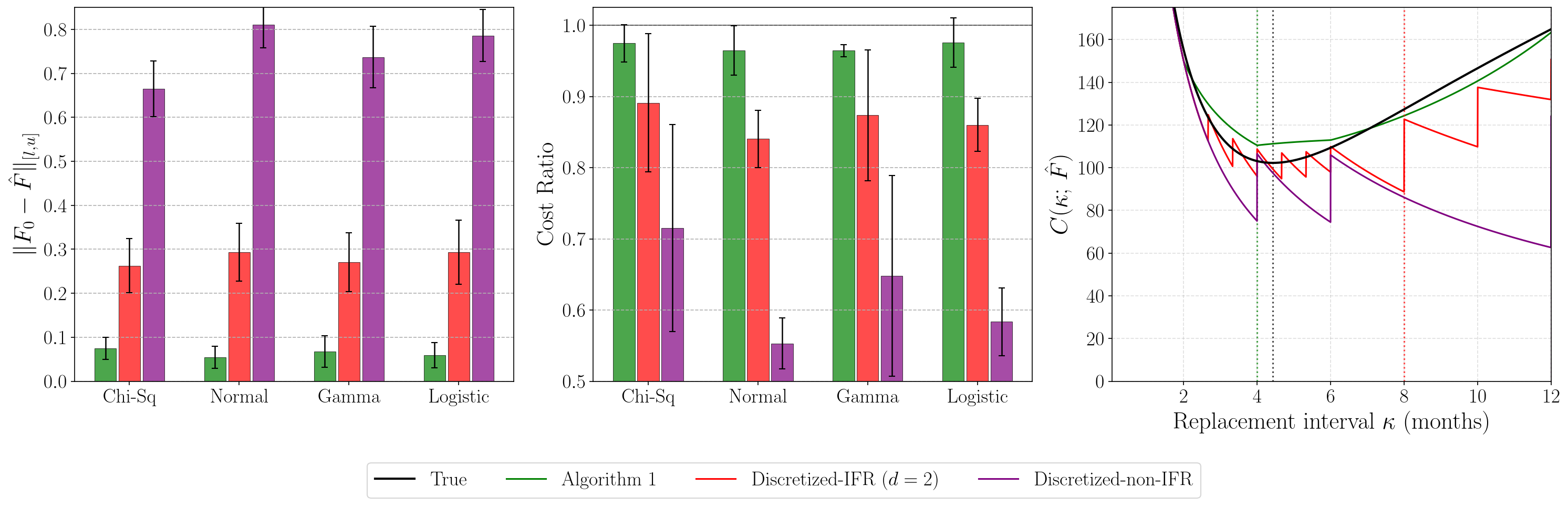}
\caption{\normalfont Preventive-maintenance case study over 150 runs with four ground-truth truncated cdfs supported on $[0,12]$: $\Ftrue\sim\text{Chi-Square}(9)$, $\Ftrue\sim\text{Normal}(9,3)$, $\Ftrue\sim\text{Gamma}(5,2)$, and $\Ftrue\sim\text{Logistic}(9,2)$. (Left) Error $\norm{\Ftrue-\Fhat}_{[l,u]}$. (Middle) Corresponding cost ratios, with the horizontal line indicating the optimal ratio of one. The bars show means and error bars show one sample standard deviation. (Right) Representative cost curves for a Chi-Square instance, with dotted vertical lines denoting the selected replacement ages.}
\label{fig:reliability_case_study}
\end{figure}

We simulate 150 runs with four ground-truth cdfs: truncated Chi-square, Normal, Gamma, and Logistic distributions, all supported on $[0,12]$. Here $a=3\varepsilon$ and $b=12\varepsilon/\mu_0$, so Proposition \ref{prop:cost-ratio} gives a cost ratio lower bound of $\frac{(1-3\varepsilon)(1-12\varepsilon/\mu_0)}{(1+3\varepsilon)(1+12\varepsilon/\mu_0)}$ whenever $\varepsilon<\min\{1/3,\mu_0/12\}$. The mean time to failure $\mu_0$ varies across the four distributions, with a smallest value of approximately $7.28$ months. Thus $\mu_0/12>1/3$ in each case, and the condition reduces to $\varepsilon<1/3$. For example, if $\varepsilon = 0.05$, the cost ratio bound is approximately 0.63. Figure \ref{fig:reliability_case_study} reports the estimation errors and cost ratios across our simulations. The results closely mirror those of the pricing case study in Figure \ref{fig:pricing_case_study}.

We conduct an additional reliability analysis that compares Algorithm \ref{alg:main} with a parametric Weibull benchmark. The Weibull family is widely used in reliability applications and provides a natural benchmark for practitioners willing to impose a parametric time-to-failure model \citep{jardine2013maintenance}. Under a Weibull specification with scale parameter $\lambda>0$ and shape parameter $\alpha>0$, the cdf is $F_0(x)=1-\exp[-(x/\lambda)^\alpha]$. The standard probability-plot transformation gives $\log[-\log(1-F_0(x))]=\alpha\log x-\alpha\log\lambda$. Thus, using the knot-level empirical cdf estimates $y_i/n_i$, we fit the Weibull benchmark by regressing\footnote{Boundary proportions $y_i/n_i\in\{0,1\}$ are replaced by half-count corrections, clipping them to $[1/(2n_i),\,1-1/(2n_i)]$. The fitted Weibull cdf is normalized to $[l,u]$; its mass above $u$ is typically below $10^{-30}$, so the adjustment is negligible.} $\log[-\log(1-y_i/n_i)]$ on $\log x_i$. If the fitted slope is $m$ and the fitted intercept is $b$, then $\hat{\alpha}=m$ and $\hat{\lambda}=\exp(-b/m)$ \citep{jardine2013maintenance}.

\begin{figure}[tbh]
    \centering

    \begin{subfigure}[t]{0.42\textwidth}
        \centering
        \includegraphics[width=\linewidth]{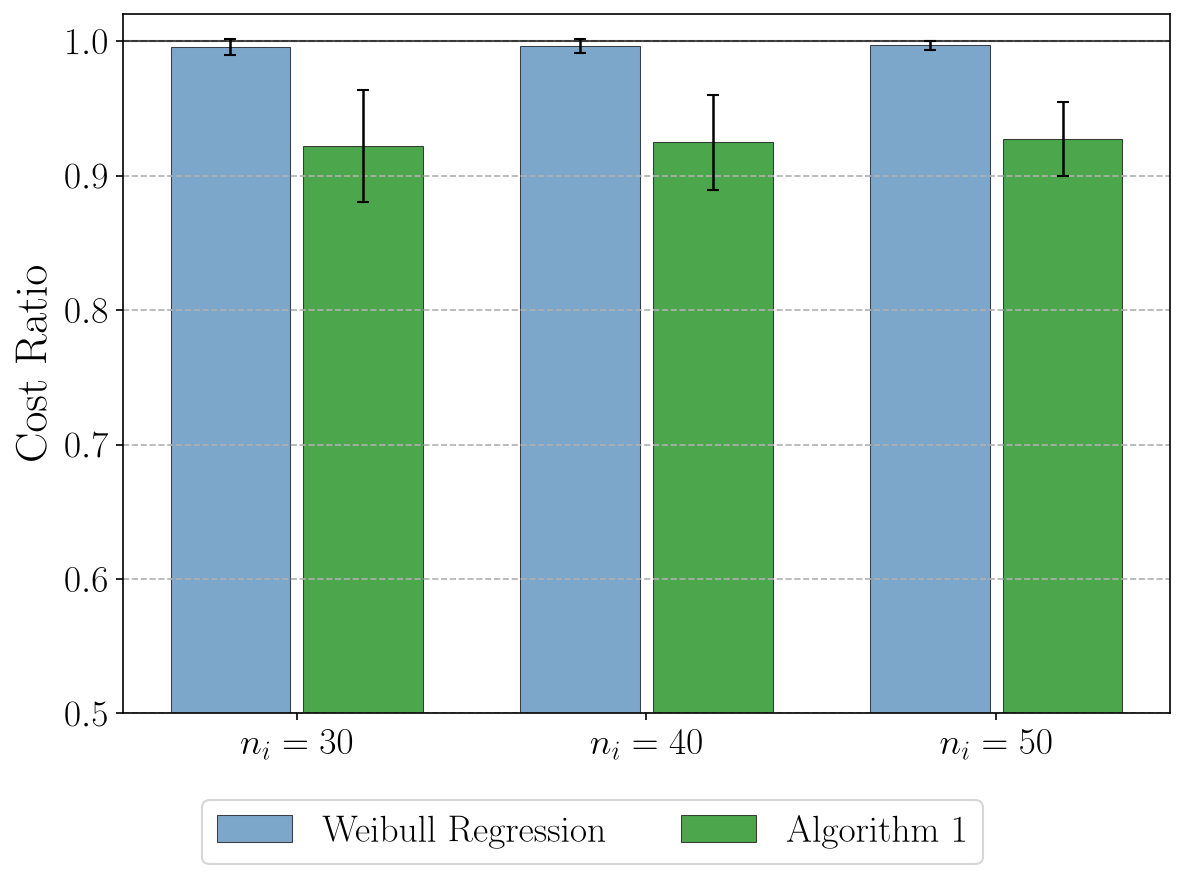}
        \caption{\small Correct specification}
        \label{fig:rel_correct_spec}
    \end{subfigure}
    \hfill
    \begin{subfigure}[t]{0.42\textwidth}
        \centering
        \includegraphics[width=\linewidth]{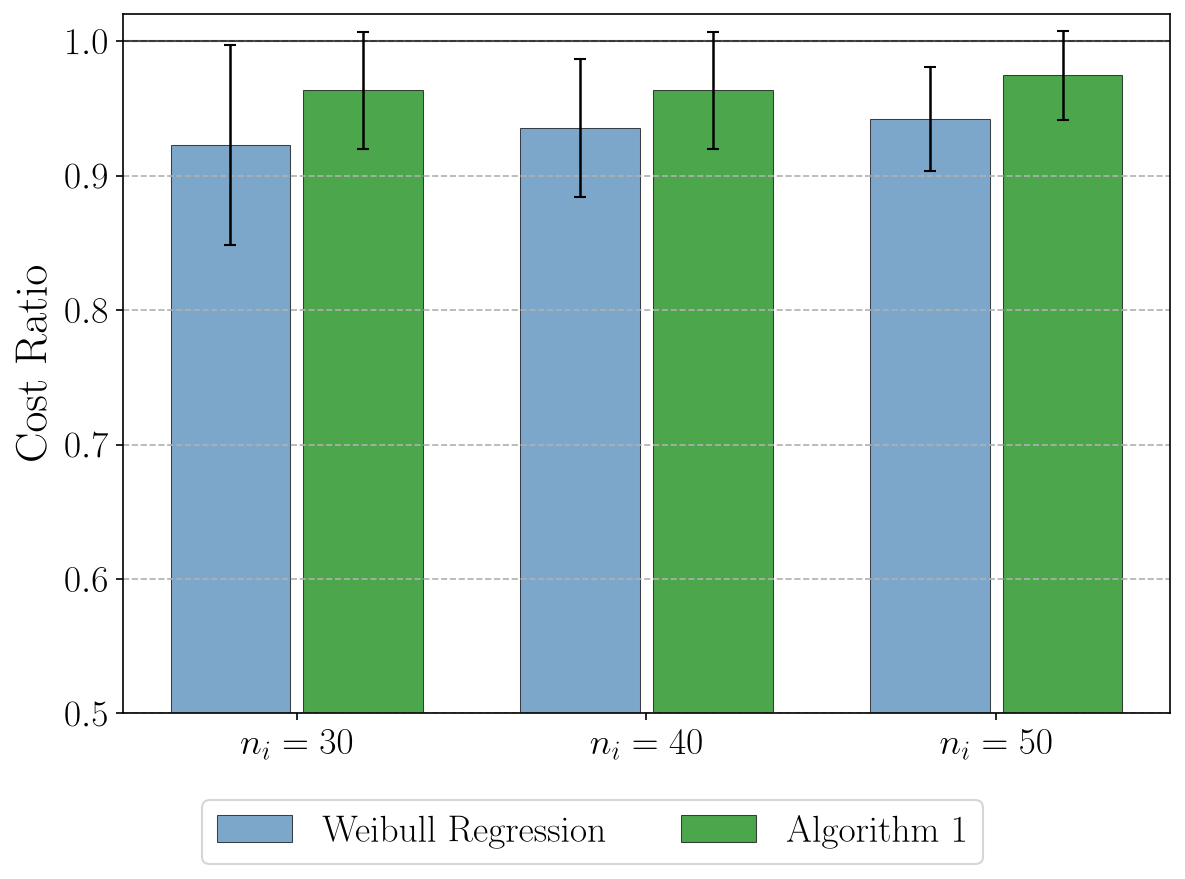}
        \caption{\small Misspecification}
        \label{fig:rel_misspec}
    \end{subfigure}
    \leaveline
    \caption{\normalfont Cost ratios for Weibull regression and Algorithm \ref{alg:main} over 150 runs, with support $[0,100]$, knots $x_1=24$, $x_2=30$, and $x_3=36$, and $c_p=600$ and $c_f=1800$. Bars show means and error bars show one sample standard deviation. (Left) Truncated Weibull$(5,30)$; (Right) Truncated two-phase hazard with $h(x)=0.005+0.02(x-28)^+$.}
    \label{fig:reliability_cost_ratio_weibull}
\end{figure}

Figure \ref{fig:reliability_cost_ratio_weibull} compares the methods when the ground-truth cdf is either a truncated Weibull$(5,30)$ or a truncated two-phase hazard distribution. In the latter, the hazard is $h(x)=0.005+0.02(x-28)^+$ on $[0,100]$: it is constant until age 28 and then increases linearly. This IFR model is not Weibull, providing a test of the parametric benchmark under misspecification. Under the Weibull ground truth, Weibull regression achieves higher mean cost ratios at all three sample sizes. Under the two-phase hazard ground truth, Algorithm \ref{alg:main} achieves higher mean cost ratios and lower run-to-run variation at every sample size. These results illustrate the benefit of exploiting a correct parametric specification and the potential cost of misspecification; in this experiment, the nonparametric IFR estimator performs better under the two-phase hazard model.

\vspace{-0.8cm}
\section{Conclusion}
\label{sec:conclusion}

In this paper, we study distribution estimation under failure-rate constraints with noisy quantile data. We show that the original maximum likelihood problem is infinite-dimensional and non-convex, and develop a tractable two-step approach that involves solving a finite-dimensional convex problem at the knots followed by interpolation to construct the full distribution.
We establish finite-sample bounds and convergence rates for our method, which provide guidance for offline data collection. Our framework accommodates several general classes of shape restrictions, including increasing failure rate, failure rate average, generalized failure rate, and new better than used. Numerical experiments and realistic case studies demonstrate that our estimator performs well in terms of both goodness-of-fit and downstream decision quality.

Several directions remain for future work. First, the reformulation approach developed here could be extended to other shape-constrained distribution classes beyond failure-rate-based restrictions. Such extensions would require identifying transformations under which the likelihood optimization problem remains convex, and the reconstruction step remains feasible. Second, adaptive data collection could be studied, in which knot locations are selected sequentially based on previously observed data rather than fixed ex ante. Such policies could improve accuracy in underexplored regions of the support or in regions most relevant to downstream decisions. Finally, although the cdf estimated by Algorithm \ref{alg:main} can be differentiated numerically, or symbolically as in our implementation, it would be valuable to develop methods that estimate the density directly while preserving the desired failure-rate structure.

\newpage

\bibliographystyle{informs2014}
\bibliography{ref}

@book{barlow1996mathematical,
  title={Mathematical theory of reliability},
  author={Barlow, Richard E and Proschan, Frank},
  year={1996},
  publisher={SIAM}
}

@article{schumaker1983shape,
  title={On shape preserving quadratic spline interpolation},
  author={Schumaker, Larry I},
  journal={SIAM Journal on Numerical Analysis},
  volume={20},
  number={4},
  pages={854--864},
  year={1983}
}

@book{bertsekas2016nonlinear,
  title={Nonlinear Programming},
  author={Bertsekas, Dimitri P.},
  edition={3rd},
  year={2016},
  publisher={Athena Scientific}
}

@article{lariviere2001selling,
  title={Selling to the newsvendor: An analysis of price-only contracts},
  author={Lariviere, Martin A and Porteus, Evan L},
  journal={Manufacturing \& Service Operations Management},
  volume={3},
  number={4},
  pages={293--305},
  year={2001},
  publisher={INFORMS}
}

@article{lariviere2006note,
  title={A note on probability distributions with increasing generalized failure rates},
  author={Lariviere, Martin A},
  journal={Operations Research},
  volume={54},
  number={3},
  pages={602--604},
  year={2006},
  publisher={INFORMS}
}

@book{ascher2011first,
  author    = {Uri M. Ascher and Chen Greif},
  title     = {A First Course in Numerical Methods},
  publisher = {SIAM},
  year      = {2011},
}

@book{jardine2013maintenance,
  title     = {Maintenance, Replacement, and Reliability: Theory and Applications},
  author    = {Jardine, A. K. S. and Tsang, A. H. C.},
  edition   = {2nd},
  year      = {2013},
  publisher = {CRC Press}
}

@article{smith1955,
    author = {Smith, W. L.},
    title = {Regenerative stochastic processes},
    journal = {Proceedings of the Royal Society of London. A. Mathematical and Physical Sciences},
    volume = {232},
    number = {1188},
    pages = {6-31},
    year = {1955},
}

@article{grenander1956,
  author    = {Ulf Grenander},
  title     = {On the theory of mortality measurement, {P}art {II}},
  journal   = {Scandinavian Actuarial Journal},
  volume    = {1956},
  number    = {2},
  pages     = {125--153},
  year      = {1956}
}

@article{marshall1965,
  author    = {Albert W. Marshall and Frank Proschan},
  title     = {Maximum Likelihood Estimation for Distributions with Monotone Failure Rate},
  journal   = {The Annals of Mathematical Statistics},
  volume    = {36},
  number    = {1},
  pages     = {69--77},
  year      = {1965}
}

@article{padgett1980,
  author    = {W. J. Padgett and L. J. Wei},
  title     = {Maximum Likelihood Estimation of a Distribution Function with Increasing Failure Rate Based on Censored Observations},
  journal   = {Biometrika},
  volume    = {67},
  number    = {2},
  pages     = {470--474},
  year      = {1980}
}

@article{bagnoli2005,
  author    = {Mark Bagnoli and Ted Bergstrom},
  title     = {Log-concave probability and its applications},
  journal   = {Economic Theory},
  volume    = {26},
  pages     = {445--469},
  year      = {2005}
}

@article{kaplan1958,
  author    = {E. L. Kaplan and Paul Meier},
  title     = {Nonparametric Estimation from Incomplete Observations},
  journal   = {Journal of the American Statistical Association},
  volume    = {53},
  number    = {282},
  pages     = {457--481},
  year      = {1958}
}

@article{prakasa1970,
  author    = {B. L. S. Prakasa Rao},
  title     = {Estimation for Distributions with Monotone Failure Rate},
  journal   = {The Annals of Mathematical Statistics},
  volume    = {41},
  number    = {2},
  pages     = {507--519},
  year      = {1970}
}

@article{caro2012clearance,
  author    = {Felipe Caro and Jérémie Gallien},
  title     = {Clearance Pricing Optimization for a Fast-Fashion Retailer},
  journal   = {Operations Research},
  volume    = {60},
  number    = {6},
  pages     = {1404--1422},
  year      = {2012}
}

@article{ferreira2015analytics,
  author    = {Kris Johnson Ferreira and Bin Hong Alex Lee and David Simchi-Levi},
  title     = {Analytics for an Online Retailer: Demand Forecasting and Price Optimization},
  journal   = {Manufacturing \& Service Operations Management},
  volume    = {18},
  number    = {1},
  pages     = {69--88},
  year      = {2016}
}

@article{besbes2020rotable,
  author    = {Omar Besbes and Adam N. Elmachtoub and Yunjie Sun},
  title     = {Pricing Analytics for Rotable Spare Parts},
  journal   = {INFORMS Journal on Applied Analytics},
  volume    = {50},
  number    = {5},
  pages     = {313--324},
  year      = {2020}
}

@article{boada2019inventory,
  author    = {Pol Boada-Collado and Victor Martínez-de-Albéniz},
  title     = {Estimating and Optimizing the Impact of Inventory on Consumer Choices in a Fashion Retail Setting},
  journal   = {Manufacturing \& Service Operations Management},
  volume    = {22},
  number    = {3},
  pages     = {582--597},
  year      = {2020}
}

@article{arslan2021sports,
  author    = {Hayri A. Arslan and Robert F. Easley and Ruxian Wang and Övünç Yılmaz},
  title     = {Data-Driven Sports Ticket Pricing for Multiple Sales Channels with Heterogeneous Customers},
  journal   = {Manufacturing \& Service Operations Management},
  volume    = {24},
  number    = {2},
  pages     = {1241--1260},
  year      = {2022}
}

@article{chen2023datadriven,
  author    = {Ningyuan Chen and Ming Hu},
  title     = {Frontiers in Service Science: Data-Driven Revenue Management: The Interplay of Data, Model, and Decisions},
  journal   = {Service Science},
  volume    = {15},
  number    = {2},
  pages     = {79--91},
  year      = {2023}
}

@article{wilder2019decisionfocused,
  author    = {Bryan Wilder and Bistra Dilkina and Milind Tambe},
  title     = {Melding the Data-Decisions Pipeline: Decision-Focused Learning for Combinatorial Optimization},
  journal   = {Proceedings of the AAAI Conference on Artificial Intelligence},
  volume    = {33},
  number    = {01},
  pages     = {1658--1665},
  year      = {2019}
}

@inproceedings{mandi2022decisionfocused,
  author    = {Jayanta Mandi and V{\'i}ctor Bucarey and Maxime Mulamba and Tias Guns},
  title     = {Decision-Focused Learning: Through the Lens of Learning to Rank},
  booktitle = {Proceedings of the 39th International Conference on Machine Learning},
  series    = {Proceedings of Machine Learning Research},
  volume    = {162},
  pages     = {14935--14947},
  year      = {2022}
}

@article{ban2019bigdata,
  author    = {Gah-Yi Ban and Cynthia Rudin},
  title     = {The Big Data Newsvendor: Practical Insights from Machine Learning},
  journal   = {Operations Research},
  volume    = {67},
  number    = {1},
  pages     = {90--108},
  year      = {2019}
}

@article{elmachtoub2021smart,
  author    = {Adam N. Elmachtoub and Paul Grigas},
  title     = {Smart ``Predict, then Optimize''},
  journal   = {Management Science},
  volume    = {68},
  number    = {1},
  pages     = {9--26},
  year      = {2022}
}

@article{bertsimas2019prescriptive,
  author    = {Dimitris Bertsimas and Nathan Kallus},
  title     = {From Predictive to Prescriptive Analytics},
  journal   = {Management Science},
  volume    = {66},
  number    = {3},
  pages     = {1025--1044},
  year      = {2020}
}

@inproceedings{fu2015randomization,
  author    = {Hu Fu and Nicole Immorlica and Brendan Lucier and Philipp Strack},
  title     = {Randomization Beats Second Price as a Prior-Independent Auction},
  booktitle = {Proceedings of the Sixteenth ACM Conference on Economics and Computation (EC '15)},
  pages     = {323},
  year      = {2015}
}

@inproceedings{huang2015samples,
  author    = {Zhiyi Huang and Yishay Mansour and Tim Roughgarden},
  title     = {Making the Most of Your Samples},
  booktitle = {Proceedings of the Sixteenth ACM Conference on Economics and Computation (EC '15)},
  pages     = {45--60},
  year      = {2015}
}

@inproceedings{daskalakis2020sdp,
  author    = {Constantinos Daskalakis and Manolis Zampetakis},
  title     = {More Revenue from Two Samples via Factor Revealing SDPs},
  booktitle = {Proceedings of the 21st ACM Conference on Economics and Computation (EC '20)},
  pages     = {257--272},
  year      = {2020}
}

@article{allouah2022pricing,
  author    = {Amine Allouah and Achraf Bahamou and Omar Besbes},
  title     = {Pricing with Samples},
  journal   = {Operations Research},
  volume    = {70},
  number    = {2},
  pages     = {1088--1104},
  year      = {2022}
}

@inproceedings{babaioff2018two,
  author    = {Moshe Babaioff and Yannai A. Gonczarowski and Yishay Mansour and Shay Moran},
  title     = {Are Two (Samples) Really Better Than One?},
  booktitle = {Proceedings of the 2018 ACM Conference on Economics and Computation (EC '18)},
  pages     = {175},
  year      = {2018},
  publisher = {Association for Computing Machinery},
  address   = {New York, NY, USA}
}

@misc{bahamou2026fast,
  author    = {Achraf Bahamou and Omar Besbes and Omar Mouchtaki},
  title     = {Fast Revenue Maximization},
  year      = {2026},
  eprint    = {2407.07316},
  archivePrefix = {arXiv},
  note      = {arXiv:2407.07316v3, revised May 2026}
}

@article{allouah2023singlepoint,
  author    = {Amine Allouah and Achraf Bahamou and Omar Besbes},
  title     = {Optimal Pricing with a Single Point},
  journal   = {Management Science},
  volume    = {69},
  number    = {10},
  pages     = {5866--5882},
  year      = {2023}
}

@misc{daei2025robust,
  author    = {Ali Daei Naby and Setareh Farajollahzadeh and Ming Hu},
  title     = {Robust One-Shot Price Experimentation},
  year      = {2025},
  note      = {Available at SSRN 4899852}
}

@book{sun2006intervalcensored,
  author    = {Sun, Jianguo},
  title     = {The Statistical Analysis of Interval-Censored Failure Time Data},
  publisher = {Springer},
  address   = {New York},
  year      = {2006}
}

@article{Turnbull1976,
 author = {Bruce W. Turnbull},
 journal = {Journal of the Royal Statistical Society. Series B (Methodological)},
 number = {3},
 pages = {290--295},
 title = {The Empirical Distribution Function with Arbitrarily Grouped, Censored and Truncated Data},
 volume = {38},
 year = {1976}
}

@incollection{wellner1992interval,
  author    = {Groeneboom, Piet and Wellner, Jon A.},
  title     = {The Interval Censoring Problem},
  booktitle = {Information Bounds and Nonparametric Maximum Likelihood Estimation},
  series    = {DMV Seminar},
  volume    = {19},
  pages     = {35--52},
  publisher = {Birkh{\"a}user Basel},
  address   = {Basel, Switzerland},
  year      = {1992}
}

@book{groeneboom2014nonparametric,
  author    = {Groeneboom, Piet and Jongbloed, Geurt},
  title     = {Nonparametric Estimation under Shape Constraints: Estimators, Algorithms and Asymptotics},
  publisher = {Cambridge University Press},
  year      = {2014}
}

@incollection{huang1997interval,
  author    = {Huang, Jian and Wellner, Jon A.},
  title     = {Interval Censored Survival Data: A Review of Recent Progress},
  booktitle = {Proceedings of the First Seattle Symposium in Biostatistics: Survival Analysis},
  series    = {Lecture Notes in Statistics},
  editor    = {Lin, D. Y. and Fleming, T. R.},
  volume    = {123},
  pages     = {123--169},
  publisher = {Springer},
  address   = {New York, NY},
  year      = {1997}
}

@article{ayer1955empirical,
  title={An empirical distribution function for sampling with incomplete information},
  author={Ayer, Miriam and Brunk, H Daniel and Ewing, George M and Reid, William T and Silverman, Edward},
  journal={The Annals of Mathematical Statistics},
  volume = {26},
  number = {4},
  pages={641--647},
  year={1955}
}

@article{dumbgen2004consistency,
  author  = {D{\"u}mbgen, Lutz and Freitag, Simone and Jongbloed, Geurt},
  title   = {Consistency of Concave Regression with an Application to Current-Status Data},
  journal = {Mathematical Methods of Statistics},
  volume  = {13},
  number  = {1},
  pages   = {69--81},
  year    = {2004}
}

@article{dumbgen2006estimating,
  author  = {D{\"u}mbgen, Lutz and Freitag-Wolf, Sandra and Jongbloed, Geurt},
  title   = {Estimating a Unimodal Distribution from Interval-Censored Data},
  journal = {Journal of the American Statistical Association},
  volume  = {101},
  number  = {475},
  pages   = {1094--1106},
  year    = {2006}
}

@article{dumbgen2014maximum,
  author  = {D{\"u}mbgen, Lutz and Rufibach, Kaspar and Schuhmacher, Dominic},
  title   = {Maximum-Likelihood Estimation of a Log-Concave Density Based on Censored Data},
  journal = {Electronic Journal of Statistics},
  volume  = {8},
  pages   = {1405--1437},
  year    = {2014}
}

@article{andersonbergman2016computing,
  author  = {Anderson-Bergman, Clifford and Yu, Yaming},
  title   = {Computing the Log Concave NPMLE for Interval Censored Data},
  journal = {Statistics and Computing},
  volume  = {26},
  pages   = {813--826},
  year    = {2016}
}

@article{chu2026nonparametric,
  author  = {Chu, Chi Wing and Ling, Hok Kan and Yuan, Chaoyu},
  title   = {Nonparametric estimation for a log-concave distribution function with interval-censored data},
  journal = {Electronic Journal of Statistics},
  volume  = {20},
  number  = {1},
  pages   = {2531--2585},
  year    = {2026}
}

@BOOK{Gallego2019,
title = {Revenue Management and Pricing Analytics},
author = {Gallego, Guillermo and Topaloglu, Huseyin},
year = {2019},
publisher = {Springer}
}

@book{tsybakov2009introduction,
  title     = {Introduction to Nonparametric Estimation},
  author    = {Tsybakov, Alexandre B.},
  year      = {2009},
  publisher = {Springer},
  address   = {New York}
}

@book{casella2002statistical,
  title     = {Statistical Inference},
  author    = {Casella, George and Berger, Roger L.},
  edition   = {2nd},
  year      = {2002},
  publisher = {Duxbury},
  address   = {Pacific Grove, CA}
}

@article{proschan1963theoretical,
  title     = {Theoretical Explanation of Observed Decreasing Failure Rate},
  author    = {Proschan, Frank},
  journal   = {Technometrics},
  volume    = {5},
  number    = {3},
  pages     = {375--383},
  year      = {1963}
}

@article{holcomb1985infant,
  title   = {An Infant Mortality and Long-Term Failure Rate Model for Electronic Equipment},
  author  = {Holcomb, Douglas P. and North, James C.},
  journal = {AT\&T Technical Journal},
  volume  = {64},
  number  = {1},
  pages   = {15--31},
  year    = {1985}
}

@article{clark2003survival,
  title   = {Survival Analysis Part I: Basic Concepts and First Analyses},
  author  = {Clark, Taane G. and Bradburn, Michael J. and Love, Sharon B. and Altman, Douglas G.},
  journal = {British Journal of Cancer},
  volume  = {89},
  number  = {2},
  pages   = {232--238},
  year    = {2003}
}

@inproceedings{leme2023description,
  author     = {Paes Leme, Renato and Sivan, Balasubramanian and Teng, Yifeng and Worah, Pratik},
  title      = {Description Complexity of Regular Distributions},
  booktitle = {Proceedings of the 24th ACM Conference on Economics and Computation},
  series     = {EC '23},
  pages      = {959},
  year       = {2023}
}

@article{diamond2016cvxpy,
  author  = {Steven Diamond and Stephen Boyd},
  title   = {{CVXPY}: {A} {P}ython-embedded modeling language for convex optimization},
  journal = {Journal of Machine Learning Research},
  year    = {2016},
  volume  = {17},
  number  = {83},
  pages   = {1--5},
}

@article{agrawal2018rewriting,
  author  = {Agrawal, Akshay and Verschueren, Robin and Diamond, Steven and Boyd, Stephen},
  title   = {A rewriting system for convex optimization problems},
  journal = {Journal of Control and Decision},
  year    = {2018},
  volume  = {5},
  number  = {1},
  pages   = {42--60},
}

@article{bertsimas2020sparse,
  author  = {Bertsimas, Dimitris and Mundru, Nishanth},
  title   = {Sparse Convex Regression},
  journal = {INFORMS Journal on Computing},
  year    = {2021},
  volume  = {33},
  number  = {1},
  pages   = {262--279}
}

@article{guntuboyina2018nonparametric,
  author  = {Guntuboyina, Adityanand and Sen, Bodhisattva},
  title   = {Nonparametric Shape-Restricted Regression},
  journal = {Statistical Science},
  year    = {2018},
  volume  = {33},
  number  = {4},
  pages   = {568--594}
}

@article{lim2012consistency,
  author  = {Lim, Eunji and Glynn, Peter W.},
  title   = {Consistency of Multidimensional Convex Regression},
  journal = {Operations Research},
  year    = {2012},
  volume  = {60},
  number  = {1},
  pages   = {196--208}
}

@article{lin2022augmented,
  author  = {Lin, Meixia and Sun, Defeng and Toh, Kim-Chuan},
  title   = {An Augmented Lagrangian Method with Constraint Generation for Shape-Constrained Convex Regression Problems},
  journal = {Mathematical Programming Computation},
  year    = {2022},
  volume  = {14},
  number = {2},
  pages   = {223--270}
  }

@misc{gurobi,
  author = {{Gurobi Optimization, LLC}},
  title = {{Gurobi Optimizer Reference Manual}},
  year = 2026
}

@article{chen2024adaptive,
  author  = {Chen, Ningyuan and Khademi, Amin},
  title   = {Adaptive Seamless Dose-Finding Trials},
  journal = {Manufacturing \& Service Operations Management},
  year    = {2024},
  volume  = {26},
  number  = {5},
  pages   = {1656--1673}
}

@article{manski2025using,
  author  = {Manski, Charles F.},
  title   = {Using Limited Trial Evidence to Credibly Choose Treatment Dosage When Efficacy and Adverse Effects Weakly Increase with Dose},
  journal = {Epidemiology},
  year    = {2025},
  volume  = {36},
  number  = {1},
  pages   = {60--65}
}

@article{schell1989increasing,
  title={An Increasing Failure Rate Approach to Low-Dose Extrapolation},
  author={Schell, Michael J. and Leysieffer, Frederick W.},
  journal={Biometrics},
  volume={45},
  number={4},
  pages={1117--1123},
  year={1989}
}

@article{babier2024learning,
  author  = {Babier, Aaron and Chan, Timothy C. Y. and Diamant, Adam and Mahmood, Rafid},
  title   = {Learning to Optimize Contextually Constrained Problems for Real-Time Decision Generation},
  journal = {Management Science},
  year    = {2025},
  volume  = {71},
  number  = {2},
  pages   = {1165--1186}
}

@article{abouee2022datadriven,
  author  = {Abouee-Mehrizi, Hossein and Mirjalili, Mahdi and Sarhangian, Vahid},
  title   = {Data-Driven Platelet Inventory Management Under Uncertainty in the Remaining Shelf Life of Units},
  journal = {Production and Operations Management},
  year    = {2022},
  volume  = {31},
  number  = {10},
  pages   = {3914--3932}
}

@article{besbes2025beyond,
  author  = {Besbes, Omar and Ma, Will and Mouchtaki, Omar},
  title   = {Beyond {IID}: Data-Driven Decision Making in Heterogeneous Environments},
  journal = {Management Science},
  year    = {2025},
  volume  = {71},
  number  = {12},
  pages   = {10538--10555}
}

@article{paul2005closure,
  author  = {Paul, Anand},
  title   = {A Note on Closure Properties of Failure Rate Distributions},
  journal = {Operations Research},
  volume  = {53},
  year    = {2005},
  number  = {4},
  pages   = {733--734}
}

@article{Birnbaum1966,
author = {Z. W. Birnbaum and J. D. Esary and A. W. Marshall},
title = {A Stochastic Characterization of Wear-out for Components and Systems},
volume = {37},
journal = {The Annals of Mathematical Statistics},
number = {4},
pages = {816 -- 825},
year = {1966}
}

\renewcommand{\theHsection}{A\arabic{section}}



\ECSwitch
\setcounter{page}{1}
\vspace{-1cm}


\section{Near-Tightness of Corollary \ref{cor:baseline_error_f}}
\label{sec:tightness-example}

In Section~\ref{sec:stochastic-estimation-error}, Corollary~\ref{cor:baseline_error_f} extends the knotwise stochastic guarantee in Proposition~\ref{prop:estimation_error_f} to the full support by adding the largest true cdf increment between consecutive knots. This bound applies to every ground-truth distribution in the estimation class and therefore provides a general consistency result. Despite this robustness, the following constructed cdf and data realization demonstrate that the bound can be nearly attained.

Consider support $[0,u]$, and let the first nine observed knots be $x_i=-\ln(1-i/10)$ for $i=1,\ldots,9$, so that $x_9=\ln 10$. Set $x_{10}=\ln 10+0.01$, $u=x_{10}+0.01$, $v=x_9+0.009$, and $b=10^{-6}$, and define $C=(-x_{10}-\ln b)/(0.001)^3$ and $d=0.01(1+3C(0.001)^2)$. Then, we can define $t_0$ as
\begin{align*}
\Ttrue(x):=
\begin{cases}
-x-C(x-v)_+^3, & x\in[0,x_{10}],\\
\ln b+d\ln\left(\dfrac{u-x}{0.01}\right), & x\in[x_{10},u),
\end{cases}
\end{align*}
where $(z)_+:=\max\{z,0\}$.

On $[0,x_{10}]$, the function $\Ttrue$ satisfies $\Ttrue'(x)=-1-3C(x-v)_+^2<0$ and $\Ttrue''(x)=-6C(x-v)_+\leq0$. Moreover, the definitions of $C$ and $d$ ensure that the logarithmic tail matches both the value and slope of the first piece at $x_{10}$ and remains decreasing and concave. Therefore, $\Ttrue$ is globally non-increasing and concave and belongs to $C^2([0,x_{10}])$. Consequently, $\Ftrue(x):=1-\exp(\Ttrue(x))$ for $x<u$, together with $\Ftrue(u):=1$, defines a valid, absolutely continuous IFR cdf with full support $[0,u]$.

Because $x_i<v$ for $i=1,\ldots,9$, the first branch of $\Ttrue$ gives $\Ftrue(x_i)=1-e^{-x_i}=\frac{i}{10},\ \forall i=1,\ldots,9$. The construction at the final knot gives $\Ftrue(x_{10})=1-b=0.999999$. Suppose that $n_i=10^8$ at every knot, with $y_i=i\times10^7$ for $i=1,\ldots,9$ and $y_{10}=99{,}999{,}900$. The empirical proportions $y_i/n_i$ therefore equal the true cdf values at all ten knots.

It is straightforward to verify that the transformed empirical proportions $\hat\tau_i:=\ln(1-y_i/n_i)$ satisfy the finite IFR constraints. Because the empirical proportions are strictly between zero and one, they uniquely maximize the unconstrained binomial likelihood at the knots. Their feasibility therefore implies that they are also the unique constrained optimal solution to \eqref{eqn:covprob}. Thus, Algorithm~\ref{alg:main} produces a valid IFR cdf that agrees exactly with $\Ftrue$ at every observed knot, although its interpolation between knots and extrapolation beyond the final knot can still incur error.

Let $r:=\frac{\ln(10^5)}{0.01}.$ The piecewise-linear and analytic-tail reconstruction in Algorithm~\ref{alg:main} is
\begin{align*}
\That(x)=
\begin{cases}
-x, & x\in[0,x_9],\\
\ln(0.1)-r(x-x_9), & x\in[x_9,x_{10}],\\
\ln b+\ln(10^5)\ln\left(\dfrac{u-x}{0.01}\right), & x\in[x_{10},u).
\end{cases}
\end{align*}

The true and fitted cdfs coincide on $[0,x_9]$. On $[x_9,v]$, write $z:=x-x_9$. The true and fitted survival functions are $0.1e^{-z}$ and $0.1e^{-rz}$, respectively, so $\Fhat(x)-\Ftrue(x)=0.1\left(e^{-z}-e^{-rz}\right)$. Differentiating shows that this expression has a unique interior maximizer at $z^*=\ln r/(r-1)$, which lies in $(0,0.009)$. Thus, $x^*:=x_9+\ln r/(r-1)\approx2.308713$. At this point, $\Ftrue(x^*)\approx0.900611$ and $\Fhat(x^*)\approx0.999914$, giving $\Fhat(x^*)-\Ftrue(x^*)\approx0.099303$. This value is the global sup-norm error. The error is zero on $[0,x_9]$. On $[v,x_{10}]$, $\hat t$ remains the chord joining $t_0(x_9)$ and $t_0(x_{10})$. Concavity of $t_0$ therefore gives $t_0(x)\geq\hat t(x)$ throughout this interval. Hence, $\Fhat\geq\Ftrue$ there, and $0\leq\Fhat(x)-\Ftrue(x)\leq1-\Ftrue(x)\leq e^{-v}=0.1e^{-0.009}\approx0.099104,
$ which is smaller than $0.099303$. Finally, on $[x_{10},u]$, both survival functions are bounded above by their common value $b=10^{-6}$ at $x_{10}$, so their difference is at most $10^{-6}$. Therefore, the realized full-support error is
\begin{align*}
\norm{\Fhat-\Ftrue}_{[0,u]}\approx0.099303.
\end{align*}

We now evaluate the terms in Corollary~\ref{cor:baseline_error_f}. The true cdf vertical increments between the first nine knots are all equal to $0.1$. The increment over $[x_9,x_{10}]$ is $0.099999$, and the terminal mass beyond $x_{10}$ is $10^{-6}$. Therefore, $\max_{i=0,\ldots,10}\left\{\Ftrue(x_{i+1})-\Ftrue(x_i)\right\}=0.1$. At confidence level $1-\delta=0.95$, the stochastic term in the bound is $\sqrt{\frac{10\ln 3+\ln(40)}{10^8}} + \sqrt{\frac{\ln(800)}{2\times10^8}} \approx0.000566$. Combining the two terms gives
\begin{align*}
\norm{\Fhat-\Ftrue}_{[0,u]}\leq0.100566.
\end{align*}
Thus, $0.099303/0.100566\approx0.987$, so the realized error is roughly $98.7\%$ of the probabilistic bound.

Finally, this ground-truth distribution has $R=\frac{1-\Ftrue(x_9)}{1-\Ftrue(x_{10})}=0.1/10^{-6}=100{,}000,$
where $R$ was defined in Section~\ref{sec:deterministic-reconstruction-error}. Consequently, the $R$-scaled stochastic term in Theorem~\ref{thm:combined_error_f} alone is approximately $56.59$, so the resulting bound is non-informative in this example, while the general bound in Corollary~\ref{cor:baseline_error_f} remains nearly tight.




\section{Proofs for Section \ref{sec:solution-framework}}
\label{apdx:proofs-sec-framework}

\ConvexProblem*

\noindent
\bold{Proof.} Consider the objective term associated with knot $x_i$, $y_i\ln(1-e^{\tau_i})+(n_i-y_i)\tau_i$. The function $z\mapsto\ln(1-e^z)$ is concave for $z<0$, so when $y_i>0$, this term is concave in $\tau_i$; assigning it the value $-\infty$ at $\tau_i=0$ preserves concavity. When $y_i=0$, the term reduces to the linear function $n_i\tau_i$ under the convention $0\log(0)=0$. Hence, the objective in \eqref{eqn:covprob} is concave in $\vecT$. Its feasible region is defined by linear equality and inequality constraints and is therefore convex. Since \eqref{eqn:covprob} maximizes a concave objective over a convex feasible region, it is a convex optimization problem. \hfill \Halmos

\leaveline
\ConvexOptUnique*

\noindent
\bold{Proof.} We temporarily consider the equivalent problem of minimizing the negative log-likelihood, $-\sum_{i=1}^k L(x_i)$, where $L(x_i):=y_i\ln(1-e^{\tau_i})+(n_i-y_i)\tau_i$ denotes the objective contribution at knot $x_i$. Suppose first that $y_k<n_k$. Because every feasible vector satisfies $0=\tau_0\geq\tau_1\geq\cdots\geq\tau_k$, any feasible sequence whose norm diverges must have $\tau_k\to-\infty$. Every negative log-likelihood contribution is nonnegative, while the contribution at the last knot tends to $+\infty$ because $n_k-y_k>0$. Hence, the negative log-likelihood is coercive over the feasible region. Under the conventions $\ln(0)=-\infty$ and $0\ln(0)=0$, the negative log-likelihood is lower semicontinuous, while the nonempty feasible region is closed because it is defined by linear equalities and inequalities. Moreover, the vector defined by $\tau_i=-(x_i-l)$ for $i=0,\ldots,k$ is feasible and has finite negative log-likelihood. The sublevel set of the negative log-likelihood at the value of this feasible vector is therefore nonempty, closed by lower semicontinuity, and bounded by coercivity, hence compact. By the compact-sublevel existence result \citep[Proposition A.8]{bertsekas2016nonlinear}, the negative log-likelihood attains its minimum on this sublevel set, which is also its global minimum because no point outside the sublevel set can improve upon the displayed feasible vector. Thus, an optimal solution exists.

Conversely, suppose that $y_k=n_k$. Given any finite feasible knot vector, decreasing $\tau_k$ preserves feasibility because it relaxes the final monotonicity and concavity constraints. It also strictly increases the final log-likelihood contribution $n_k\ln(1-e^{\tau_k})$, which approaches zero as $\tau_k\to-\infty$. Thus, no finite feasible knot vector can be optimal, and the supremum is not attained. Therefore, an optimal solution to \eqref{eqn:covprob} exists if and only if $y_k<n_k$.

Now suppose that $y_k<n_k$ and $y_i>0$ for every $i=1,\ldots,k$. The negative log-likelihood is strictly convex in $(\tau_1,\ldots,\tau_k)$ because $z\mapsto-\ln(1-e^z)$ is strictly convex for $z<0$ and the remaining terms are linear. Since the feasible region is convex, the minimizing knot vector is unique.

When $y_i=0$ at one or more knots, the corresponding objective terms are linear, and nonuniqueness can occur. For example, let $l=0$, $u>3$, $x_i=i$ for $i=1,2,3$, $n_i=1$ for all $i$, and $(y_1,y_2,y_3)=(0,1,0)$. For every feasible knot vector, concavity implies $\tau_1+\tau_3\leq2\tau_2$, so its log-likelihood is at most $2\tau_2+\ln(1-e^{\tau_2})$. This expression is maximized at $q:=\ln(2/3)$. Both knot vectors $(0,q/2,q,3q/2)$ and $(0,0,q,2q)$ are feasible and attain this bound. Hence, an optimal knot vector need not be unique when some $y_i=0$.
\hfill \Halmos



\leaveline
\KnotValueEquivalence*

\noindent
\bold{Proof.} We prove the result by establishing both inequalities between the two supremum values. First, consider any finite feasible solution $\vecT$ to \eqref{eqn:covprob}. Piecewise-linear interpolation of its knot values produces a non-increasing, concave function on $[l,x_k]$. Appending a logarithmic tail on $[x_k,u)$ whose initial slope preserves concavity and that tends to $-\infty$ as $x\to u$ produces a full transformed function $t$. The corresponding function $F=1-e^t$ is an absolutely continuous IFR cdf feasible for \eqref{eqn:originalprob}. Because it agrees with $\vecT$ at every observed knot, it has the same objective value. Therefore, the supremum value of \eqref{eqn:originalprob} is at least that of \eqref{eqn:covprob}.

For the reverse inequality, consider any feasible $F$ for \eqref{eqn:originalprob} having finite objective value, and let $t(x)=\ln(1-F(x))$. If $t(x_i)$ is finite at every observed knot, then $(t(x_0),\ldots,t(x_k))$ is feasible for \eqref{eqn:covprob} and has the same objective value. Otherwise, let $j$ be the largest index such that $t(x_j)>-\infty$, with $j:=0$ if $t(x_i)=-\infty$ for every observed knot. Since $t$ is non-increasing, $t(x_i)=-\infty$ for every $i=j+1,\ldots,k$. Finiteness of the objective requires $y_i=n_i$ at each of these knots.

For each sufficiently large $m>0$, define a finite knot vector $\vecT^{(m)}$ by
\begin{align*}
\tau_i^{(m)}:=\begin{cases}t(x_i), & i=0,\ldots,j,\\ t(x_j)-m(x_i-x_j), & i=j+1,\ldots,k.\end{cases}
\end{align*}
When $j\geq1$, choose $m$ sufficiently large that
\begin{align*}
-m\leq\frac{t(x_j)-t(x_{j-1})}{x_j-x_{j-1}};
\end{align*}
when $j=0$, any $m>0$ is valid. The retained knot values satisfy the finite monotonicity and concavity constraints, the appended knot values have constant slope $-m$, and the displayed inequality preserves concavity at $x_j$. Hence, $\vecT^{(m)}$ is feasible for \eqref{eqn:covprob}. As $m\to\infty$, each appended value tends to $-\infty$, so its all-success likelihood contribution $n_i\ln(1-e^{\tau_i^{(m)}})$ tends to zero, which is the corresponding contribution under $F(x_i)=1$. Thus, for every $\epsilon>0$, a sufficiently large $m$ gives a feasible solution to \eqref{eqn:covprob} whose objective value is within $\epsilon$ of that of $F$. Taking suprema shows that the supremum value of \eqref{eqn:covprob} is at least that of \eqref{eqn:originalprob}.

Combining the two inequalities proves that \eqref{eqn:covprob} and \eqref{eqn:originalprob} have the same supremum value.
\hfill \Halmos
\leaveline
\ReconstructionCoincides*

\bold{Proof.} By construction, $\That$ is non-increasing and concave, satisfies the endpoint conditions, and yields an absolutely continuous cdf $\Fhat$. Thus, every cdf returned by Algorithm \ref{alg:main} is feasible for \eqref{eqn:originalprob}.

Suppose first that $y_k<n_k$. By Lemma \ref{lem:opt_unique}, \eqref{eqn:covprob} admits an optimal knot vector, and Algorithm \ref{alg:main} may use any such vector. Because $\Fhat$ agrees with this vector at every observed knot, it attains the supremum value of \eqref{eqn:covprob}, which equals that of \eqref{eqn:originalprob} by Lemma \ref{lem:knot_value_equivalence}. Hence, $\Fhat$ is optimal for \eqref{eqn:originalprob}.

Now suppose that $y_k=n_k$, and let $j$ be defined as in Algorithm \ref{alg:main}. If $j=0$, every observation is a success, so the log-likelihood is at most zero, while Algorithm \ref{alg:main} attains zero by setting $\Fhat(x_i)=1$ for every $i=1,\ldots,k$. If $j\geq1$, every finite feasible solution to \eqref{eqn:covprob} restricts to a feasible solution of the reduced problem on $x_1,\ldots,x_j$, while each terminal all-success contribution is non-positive. Therefore, the supremum value of \eqref{eqn:covprob} is no greater than the optimal value of the reduced problem. Algorithm \ref{alg:main} attains the latter value by using an optimal reduced knot vector and setting $\Fhat(x_i)=1$ for $i=j+1,\ldots,k$. Since $\Fhat$ is feasible for \eqref{eqn:originalprob} and Lemma \ref{lem:knot_value_equivalence} equates its supremum value with that of \eqref{eqn:covprob}, all these bounds hold with equality. Thus, $\Fhat$ is optimal.

Finally, because Algorithm \ref{alg:main} returns an optimal cdf for every realization of the data, \eqref{eqn:originalprob} always admits an optimal solution.\hfill \Halmos

\section{Proofs for Section \ref{sec:theory}}
\label{apdx:proofs-sec-theory}

\leaveline
\EstimationErrorFspace*

\noindent\bold{Proof.} For each knot $i=1,\ldots,k$, let $p_i:=\Ftrue(x_i)$, $\hat p_i:=\Fhat(x_i)$, and $\bar p_i:=y_i/n_i$. For Bernoulli probabilities $q$ and $r$, write $\operatorname{kl}(q,r):=q\ln(q/r)+(1-q)\ln((1-q)/(1-r))$, with the usual boundary conventions, and define $Z_i:=n_i\operatorname{kl}(\bar p_i,p_i)$. Thus, $Z_i$ measures the empirical log-likelihood loss from replacing the empirical proportion $\bar p_i$ by the true cdf value $p_i$.

A standard binomial Chernoff bound gives $\bP(Z_i>z)\leq2e^{-z}$ for every $z\geq0$. Because $Z_i\geq0$, integration of its tail yields
\begin{align*}
\bE\left[e^{Z_i/2}\right]&=1+\frac{1}{2}\int_0^\infty e^{z/2}\bP(Z_i>z)\,dz \leq1+\int_0^\infty e^{-z/2}\,dz=3.
\end{align*}
The observations are independent across knots, and hence
\begin{align*}
\bE\left[\exp\left(\frac{1}{2}\sum_{i=1}^k Z_i\right)\right]\leq3^k.
\end{align*}
Markov's inequality therefore implies
\begin{align*}
\bP\left(\sum_{i=1}^k Z_i>2\left\{k\ln3+\ln(2/\delta)\right\}\right)\leq\frac{\delta}{2}.
\end{align*}
Consequently, with probability at least $1-\delta/2$, $\sum_{i=1}^k Z_i\leq2\{k\ln3+\ln(2/\delta)\}$.

We next relate this empirical likelihood loss to the fitted knot values. By Theorem~\ref{thm:optimal-solution}, the cdf $\Fhat$ returned by Algorithm~\ref{alg:main} maximizes the likelihood, including in the terminal-all-success branch. Since $\Ftrue$ is feasible, the likelihood of $\Fhat$ is at least that of $\Ftrue$. Rewriting the Bernoulli log-likelihood in terms of Kullback--Leibler divergence therefore gives
\begin{align*}
\sum_{i=1}^k n_i\operatorname{kl}(\bar p_i,\hat p_i)\leq\sum_{i=1}^k n_i\operatorname{kl}(\bar p_i,p_i)=\sum_{i=1}^k Z_i.
\end{align*}
This comparison remains valid at terminal all-success knots: there $\bar p_i=\hat p_i=1$ and $\operatorname{kl}(1,1)=0$ under the boundary conventions used in the formulation.

Pinsker's inequality gives $\operatorname{kl}(q,r)\geq2(q-r)^2$ for Bernoulli probabilities. Thus, on the preceding likelihood event, for every $i=1,\ldots,k$,
\begin{align*}
\abs{\hat p_i-\bar p_i}&\leq\sqrt{\frac{1}{2n_i}\sum_{j=1}^k n_j\operatorname{kl}(\bar p_j,\hat p_j)} \leq\sqrt{\frac{1}{2n_i}\sum_{j=1}^k Z_j} \leq\sqrt{\frac{k\ln3+\ln(2/\delta)}{n_i}}.
\end{align*}
This controls the adjustment from the empirical proportions to the likelihood-maximizing fitted knot values.

It remains to control ordinary binomial sampling error. Hoeffding's inequality gives $\bP\bigl(\abs{\bar p_i-p_i}>\sqrt{\ln(4k/\delta)/(2n_i)}\bigr)\leq\delta/(2k)$. A union bound therefore shows that, with probability at least $1-\delta/2$, $\abs{\bar p_i-p_i}\leq\sqrt{\ln(4k/\delta)/(2n_i)}$ simultaneously for every $i$.

Intersecting the likelihood and sampling events produces an event of probability at least $1-\delta$. On this event, the triangle inequality gives, for every $i=1,\ldots,k$,
\begin{align*}
\abs{\hat p_i-p_i}&\leq\abs{\hat p_i-\bar p_i}+\abs{\bar p_i-p_i} \leq\sqrt{\frac{k\ln3+\ln(2/\delta)}{n_i}}+\sqrt{\frac{\ln(4k/\delta)}{2n_i}}.
\end{align*}
Taking the maximum over the knots proves the result. \hfill \Halmos

\leaveline
\BaselineErrorFspace*

\noindent\bold{Proof.} Let $\varepsilon:=\max_{i=1,\ldots,k}\abs{\Fhat(x_i)-\Ftrue(x_i)}$. At the support endpoints, $\Fhat(x_0)=\Ftrue(x_0)=0$ and $\Fhat(x_{k+1})=\Ftrue(x_{k+1})=1$, so the corresponding endpoint errors are also bounded by $\varepsilon$.

Fix $i\in\{0,\ldots,k\}$ and $x\in[x_i,x_{i+1}]$. Because both $\Fhat$ and $\Ftrue$ are non-decreasing,
\begin{align*}
\Fhat(x)-\Ftrue(x)&\leq\Fhat(x_{i+1})-\Ftrue(x_i)\\
&=\Fhat(x_{i+1})-\Ftrue(x_{i+1})+\Ftrue(x_{i+1})-\Ftrue(x_i)\\
&\leq\varepsilon+\Ftrue(x_{i+1})-\Ftrue(x_i),
\end{align*}
and, in the other direction,
\begin{align*}
\Ftrue(x)-\Fhat(x)&\leq\Ftrue(x_{i+1})-\Fhat(x_i)\\
&=\Ftrue(x_{i+1})-\Ftrue(x_i)+\Ftrue(x_i)-\Fhat(x_i)\\
&\leq\varepsilon+\Ftrue(x_{i+1})-\Ftrue(x_i).
\end{align*}
It follows that $\abs{\Fhat(x)-\Ftrue(x)}\leq\varepsilon+\Ftrue(x_{i+1})-\Ftrue(x_i).$ Taking the supremum over $x\in[l,u]$ and then applying Proposition~\ref{prop:estimation_error_f} proves the first displayed bound in the corollary.

If $\norm{f_0}_{[l,u]}<\infty$, then, for $i=0,\ldots,k-1$, absolute continuity gives
\begin{align*}
\Ftrue(x_{i+1})-\Ftrue(x_i)=\int_{x_i}^{x_{i+1}}f_0(z)\,dz\leq\norm{f_0}_{[l,u]}(x_{i+1}-x_i)\leq\norm{f_0}_{[l,u]}\Delta.
\end{align*}
The terminal increment is $\Ftrue(u)-\Ftrue(x_k)=1-\Ftrue(x_k)$. Taking the maximum over all intervals and substituting into the preceding bound proves the bounded-density specialization. \hfill \Halmos

\leaveline
\ReconstructionStability*

\noindent\bold{Proof.} Let $\varepsilon:=\max_{i=1,\ldots,k}\abs{\Fhat(x_i)-\Ftrue(x_i)}$. Since $\Fhat(l)=\Ftrue(l)=0$, the same bound also holds at $x_0=l$. We first consider any interval $[x_i,x_{i+1}]$ on which Algorithm~\ref{alg:main} uses piecewise-linear interpolation in transformed space.

Write $a:=1-\Ftrue(x_i)$, $b:=1-\Ftrue(x_{i+1})$, $A:=1-\Fhat(x_i)$, and $B:=1-\Fhat(x_{i+1})$. The full-support assumption and $x_{i+1}\leq x_k<u$ imply $a\geq b>0$. For $x=(1-\theta)x_i+\theta x_{i+1}$ with $\theta\in[0,1]$, piecewise linearity in transformed space is equivalent to geometric interpolation of survival:
\begin{align*}
1-\Ftilde(x)=a^{1-\theta}b^\theta,\qquad 1-\Fhat(x)=A^{1-\theta}B^\theta.
\end{align*}

The knotwise error bound gives $A\leq a+\varepsilon\leq(1+\varepsilon/b)a$ and $B\leq b+\varepsilon=(1+\varepsilon/b)b$, where the first inequality uses $a\geq b$. Monotonicity and homogeneity of the geometric mean then give
\begin{align*}
1-\Fhat(x)\leq\left(1+\frac{\varepsilon}{b}\right)a^{1-\theta}b^\theta=\left(1+\frac{\varepsilon}{b}\right)(1-\Ftilde(x)).
\end{align*}

Case (i): $\varepsilon<b$. We similarly have $A\geq a-\varepsilon\geq(1-\varepsilon/b)a$ and $B\geq b-\varepsilon=(1-\varepsilon/b)b$, and therefore $1-\Fhat(x)\geq\left(1-\frac{\varepsilon}{b}\right)(1-\Ftilde(x)).$ Combining the preceding two comparisons and using $1-\Ftilde(x)\leq a$ gives $\abs{\Fhat(x)-\Ftilde(x)}\leq\frac{\varepsilon}{b}(1-\Ftilde(x))\leq\frac{a}{b}\varepsilon\leq R\varepsilon.$

Case (ii): $\varepsilon\ge b$. The upper comparison still gives $\Ftilde(x)-\Fhat(x)=\bigl(1-\Fhat(x)\bigr)-\bigl(1-\Ftilde(x)\bigr)\leq\frac{\varepsilon}{b}(1-\Ftilde(x))\leq R\varepsilon.$ For the opposite direction, the fitted survival is nonnegative, so $\Fhat(x)-\Ftilde(x)=\bigl(1-\Ftilde(x)\bigr)-\bigl(1-\Fhat(x)\bigr)\leq1-\Ftilde(x)\leq a\leq\frac{a}{b}\varepsilon\leq R\varepsilon.$

Thus, $\abs{\Fhat(x)-\Ftilde(x)}\leq R\varepsilon$ on every interval where the fitted reconstruction is piecewise-linear in transformed space. If $y_k<n_k$, these intervals cover all of $[l,x_k]$, and the proof is complete. 

Suppose instead that Algorithm~\ref{alg:main} enters the terminal-all-success branch, and let $x_j$ be the last knot retained in the finite optimization. The argument below also covers $j=0$, in which case $x_j=l$ and $\Fhat(x_j)=\Ftrue(x_j)=0$. Write $a:=1-\Ftrue(x_j)$, $b:=1-\Ftrue(x_{j+1})$, and $A:=1-\Fhat(x_j)$. Because $\Fhat(x_{j+1})=1$, we have $b=\abs{\Fhat(x_{j+1})-\Ftrue(x_{j+1})}\leq\varepsilon$.

On $[x_j,x_{j+1}]$, the fitted survival decreases continuously from $A$ to zero, whereas the oracle survival lies between $b$ and $a$. Since $a/b\leq R$ and $A\leq a+\varepsilon$, the two possible error directions satisfy
\begin{align*}
\Fhat(x)-\Ftilde(x)&=\bigl(1-\Ftilde(x)\bigr)-\bigl(1-\Fhat(x)\bigr)\leq a\leq Rb\leq R\varepsilon,\\
\Ftilde(x)-\Fhat(x)&=\bigl(1-\Fhat(x)\bigr)-\bigl(1-\Ftilde(x)\bigr)\leq A-b\\
&\leq a-b+\varepsilon\leq(R-1)b+\varepsilon\leq R\varepsilon.
\end{align*}
For $x\in[x_{j+1},x_k]$, the fitted survival is zero, while the oracle survival is no larger than its value $b$ at $x_{j+1}$. Hence, $\abs{\Fhat(x)-\Ftilde(x)}\leq b\leq\varepsilon\leq R\varepsilon$ on the remainder of the terminal-all-success region. Taking the supremum over $[l,x_k]$ proves the lemma. \hfill \Halmos

\leaveline
\InterpolationErrorFspace*

\noindent\bold{Proof.} Fix an interval $[x_i,x_{i+1}]$, where $i\in\{0,\ldots,k-1\}$. Since $\Ttrue$ is concave, it lies above the chord joining its endpoint values. This chord is precisely $\Ttilde$ on $[x_i,x_{i+1}]$, so $\Ttrue(x)-\Ttilde(x)\geq0$ throughout the interval. Because $t\mapsto1-e^t$ is decreasing, it follows immediately that $\Ftilde(x)\geq\Ftrue(x)$.

Let $r(x):=\Ttrue(x)-\Ttilde(x)\geq0$. Since $\Ttilde(x)=\Ttrue(x)-r(x)$,
\begin{align*}
\Ftilde(x)-\Ftrue(x)&=\exp(\Ttrue(x))-\exp(\Ttilde(x))\\
&=\bigl(1-\Ftrue(x)\bigr)\left(1-e^{-r(x)}\right)\\
&\leq\bigl(1-\Ftrue(x)\bigr)r(x),
\end{align*}
where the last inequality uses $1-e^{-z}\leq z$ for $z\geq0$.

The standard piecewise-linear interpolation remainder \citep[Chapter 11]{ascher2011first} gives
\begin{align*}
0\leq r(x)\leq\frac{(x_{i+1}-x_i)^2}{8}\norm{\Ttrue''}_{[x_i,x_{i+1}]}.
\end{align*}
Moreover, $1-\Ftrue$ is non-increasing, so $1-\Ftrue(x)\leq1-\Ftrue(x_i)$ on this interval. Consequently,
\begin{align*}
0\leq\Ftilde(x)-\Ftrue(x)\leq\frac{(x_{i+1}-x_i)^2}{8}\bigl(1-\Ftrue(x_i)\bigr)\norm{\Ttrue''}_{[x_i,x_{i+1}]}.
\end{align*}

For every $z\in[x_i,x_{i+1}]$, monotonicity of survival and the definition of $R$ imply $1-\Ftrue(x_i)\leq R\bigl(1-\Ftrue(z)\bigr).$ Multiplying this pointwise inequality by $\abs{\Ttrue''(z)}$ and taking the supremum over $z\in[x_i,x_{i+1}]$ yields $\bigl(1-\Ftrue(x_i)\bigr)\norm{\Ttrue''}_{[x_i,x_{i+1}]}\leq R\norm{(1-\Ftrue)\Ttrue''}_{[x_i,x_{i+1}]}.$ Since $x_{i+1}-x_i\leq\Delta$, taking the maximum over $i=0,\ldots,k-1$ proves
\begin{align*}
\norm{\Ftilde-\Ftrue}_{\lxk}\leq\frac{R\Delta^2}{8}\norm{(1-\Ftrue)\Ttrue''}_{\lxk}.
\end{align*} \hfill \Halmos

\leaveline
\CombinedErrorFspace*

\noindent\bold{Proof.} Let $\varepsilon:=\max_{i=1,\ldots,k}\abs{\Fhat(x_i)-\Ftrue(x_i)}$. On the observed region, inserting the oracle reconstruction and applying the triangle inequality gives
\begin{align*}
\norm{\Fhat-\Ftrue}_{\lxk}&\leq\norm{\Fhat-\Ftilde}_{\lxk}+\norm{\Ftilde-\Ftrue}_{\lxk}\\
&\leq R\varepsilon+\frac{R\Delta^2}{8}\norm{(1-\Ftrue)\Ttrue''}_{\lxk},
\end{align*}
where the second line follows from Lemmas~\ref{lem:reconstruction_stability} and~\ref{lem:interpolation_error_f}. On the event in Proposition~\ref{prop:estimation_error_f}, substituting its bound for $\varepsilon$ proves the first inequality in the theorem.

It remains to control the terminal region. Let $\hat S:=1-\Fhat$ and $S_0:=1-\Ftrue$. Both survival functions are non-increasing and nonnegative. Therefore, for every $x\in[x_k,u]$,
\begin{align*}
\abs{\Fhat(x)-\Ftrue(x)}&=\abs{\hat S(x)-S_0(x)}\\
&\leq\max\{\hat S(x),S_0(x)\}\\
&\leq\max\{\hat S(x_k),S_0(x_k)\}\\
&\leq S_0(x_k)+\abs{\hat S(x_k)-S_0(x_k)}\\
&=1-\Ftrue(x_k)+\abs{\Fhat(x_k)-\Ftrue(x_k)}\\
&\leq1-\Ftrue(x_k)+\varepsilon \\
& \leq1-\Ftrue(x_k)+R\varepsilon.
\end{align*}
Thus, we have $\norm{\Fhat-\Ftrue}_{(x_k,u]} \le R\varepsilon + 1-\Ftrue(x_k)$. Now combine this with the observed-region bound:
\begin{align*}
\norm{\Fhat-\Ftrue}_{[l,u]}\leq R\varepsilon+\max\left\{\frac{R\Delta^2}{8}\norm{(1-\Ftrue)\Ttrue''}_{\lxk},1-\Ftrue(x_k)\right\}.
\end{align*}
Finally, substituting the knotwise bound from Proposition~\ref{prop:estimation_error_f} on the same probability-$1-\delta$ event proves the full-support inequality. No additional probability loss is incurred because the reconstruction and interpolation bounds are deterministic. \hfill \Halmos

\section{Proofs for Section \ref{sec:case-study}}
\label{apdx:proofs-sec-case-study}

\leaveline
\PricingRevenueRatio*

\noindent\bold{Proof.} For any $p\in[l,u]$, $\abs{\mathcal R(p;\Fhat)-\mathcal R(p;\Ftrue)}\leq p\varepsilon$. Since $p^*$ maximizes true revenue and $\hat p$ maximizes estimated revenue, we have
\begin{align*}
0\leq\mathcal R(p^*;\Ftrue)-\mathcal R(\hat p;\Ftrue)
&\leq\bigl[\mathcal R(p^*;\Ftrue)-\mathcal R(p^*;\Fhat)\bigr]+\bigl[\mathcal R(\hat p;\Fhat)-\mathcal R(\hat p;\Ftrue)\bigr]\leq(p^*+\hat p)\varepsilon.
\end{align*}
Dividing by $\mathcal R(p^*;\Ftrue)>0$ and using the nonnegativity of the revenue ratio gives the first inequality. The second follows from $\mathcal R(p^*;\Ftrue)\geq\mathcal R(l;\Ftrue)=l$ and $p^*+\hat p\leq2u$. \hfill \Halmos

\leaveline
\PreventiveMaintenanceCostRatio*

\noindent\bold{Proof.} Write $N_F(\kappa):=c_p+(c_f-c_p)F(\kappa)$ and $A_F(\kappa):=\int_0^\kappa(1-F(t))\,dt$, so $C(\kappa;F)=N_F(\kappa)/A_F(\kappa)$. For every $\kappa\in(0,u]$, the definition of $\varepsilon$ and $N_{\Ftrue}(\kappa)\geq c_p$ give
\begin{align*}
\abs{N_{\Fhat}(\kappa)-N_{\Ftrue}(\kappa)}&=(c_f-c_p)\abs{\Fhat(\kappa)-\Ftrue(\kappa)}\leq(c_f-c_p)\varepsilon=a c_p\leq aN_{\Ftrue}(\kappa).
\end{align*}
Because $1-\Ftrue$ is non-increasing, its average over $[0,\kappa]$ is at least its average over $[0,u]$. Thus $A_{\Ftrue}(\kappa)/\kappa\geq A_{\Ftrue}(u)/u=\mu_0/u$, and
\begin{align*}
\abs{A_{\Fhat}(\kappa)-A_{\Ftrue}(\kappa)}&\leq\int_0^\kappa\abs{\Fhat(t)-\Ftrue(t)}\,dt\leq\kappa\varepsilon\leq bA_{\Ftrue}(\kappa).
\end{align*}

Hence $N_{\Fhat}(\kappa)$ lies between $(1-a)N_{\Ftrue}(\kappa)$ and $(1+a)N_{\Ftrue}(\kappa)$, while $A_{\Fhat}(\kappa)$ lies between $(1-b)A_{\Ftrue}(\kappa)$ and $(1+b)A_{\Ftrue}(\kappa)$. Since $a,b<1$, all relevant denominators are positive, and therefore
\begin{align*}
\frac{1-a}{1+b}C(\kappa;\Ftrue)&\leq C(\kappa;\Fhat)\leq\frac{1+a}{1-b}C(\kappa;\Ftrue).
\end{align*}

Apply the lower bound at $\hat\kappa$, use $C(\hat\kappa;\Fhat)\leq C(\kappa^*;\Fhat)$, and then apply the upper bound at $\kappa^*$:
\begin{align*}
C(\hat\kappa;\Ftrue)&\leq\frac{1+b}{1-a}C(\hat\kappa;\Fhat)\leq\frac{1+b}{1-a}C(\kappa^*;\Fhat)\leq\frac{(1+a)(1+b)}{(1-a)(1-b)}C(\kappa^*;\Ftrue).
\end{align*}
Both true costs are positive, so rearranging the last inequality gives the desired result.
\hfill \Halmos

\end{document}